\documentclass[journal]{IEEEtran}
\IEEEoverridecommandlockouts

\usepackage{graphicx}
\usepackage{amsmath}

\usepackage{amssymb}
\usepackage{amsfonts}
\usepackage[font=small]{caption}
\usepackage{comment}
\usepackage[normalem]{ulem}
\usepackage{makecell}
\usepackage{multirow}
\usepackage{caption}
\usepackage{subcaption}
\usepackage{diagbox}
\usepackage{url}

\AtBeginDocument{%
  }

\usepackage{tikzpagenodes,etoolbox}
\usetikzlibrary{calc}
\usepackage[contents={}]{background}
\AddEverypageHook{%
\ifnumequal{\thepage}{1}{%
 \tikz[remember picture,overlay]{%
     \node[draw,
     minimum width=\textwidth,
     text width=\textwidth,
     font=\footnotesize
     ]
     at ($(current page header area) - (0,5pt)$)
     {%
     This paper has been accepted for publication in {\em IEEE Transactions on Network and Service Management}. This is the authors' accepted version of the manuscript. The final version published by IEEE is “A. Calagna, Y. Yu, P. Giaccone and C. F. Chiasserini, “MOSE: A Novel Orchestration Framework for Stateful Microservice Migration at the Edge," in {\em IEEE Transactions on Network and Service Management}, doi: 10.1109/TNSM.2025.3579051."
     };
 }%
}{}%
}

\begin{document}

\title{MOSE: A Novel Orchestration Framework for Stateful Microservice Migration at the Edge
\thanks{The work was supported by the EC through Grant No.\,101139266 (6G-INTENSE  project), Grant No.\,101095363 (ADROIT6G project) and Grant No.\,101095871 (TRIALSNET project). The work of Y. Yu was supported by Leonardo S.p.A.}}

\author{\IEEEauthorblockN{Antonio Calagna\IEEEauthorrefmark{1},~\IEEEmembership{Student Member,~IEEE}, Yenchia Yu\IEEEauthorrefmark{1},~\IEEEmembership{Student Member,~IEEE}, Paolo~Giaccone\IEEEauthorrefmark{1},~\IEEEmembership{Senior~Member,~IEEE}, Carla Fabiana Chiasserini\IEEEauthorrefmark{1}\IEEEauthorrefmark{2}\IEEEauthorrefmark{3},~\IEEEmembership{Fellow,~IEEE} \\
\IEEEauthorblockA{\IEEEauthorrefmark{1}Politecnico di Torino, Italy, \IEEEauthorrefmark{2}CNIT, Italy, \IEEEauthorrefmark{3}Chalmers University, Sweden}}

}

\maketitle

\begin{abstract}
Stateful migration has emerged as the dominant technology to support microservice mobility at the network edge while ensuring a satisfying experience to mobile end users.
This work addresses two pivotal challenges, namely, the implementation and the orchestration of the migration process. We first introduce a novel framework that efficiently implements stateful migration and effectively orchestrates the migration process by fulfilling both network and application KPI targets. Through experimental validation using realistic microservices, we then show that our solution (i) greatly improves migration performance, yielding up to 77\% decrease of the migration downtime with respect to the state of the art, and (ii)  successfully addresses the strict user QoE requirements of critical scenarios featuring latency-sensitive microservices. 
Further, we consider two practical use cases, featuring, respectively, a UAV autopilot microservice and a multi-object tracking task, and demonstrate how our framework outperforms current state-of-the-art approaches in configuring the migration process and in meeting  KPI targets.

\end{abstract}


\section{Introduction}\label{work05:sec:introduction}
In recent years, edge computing has been acknowledged as the state-of-the-art paradigm to bring applications, computational capabilities, and storage facilities closer to the end users. To fully exploit the benefits of edge computing architectures, applications are increasingly designed in the form of MicroServices (MSs) chains, taking advantage of the lightweight container virtualization technology.
Concurrently, due to the steady growth of mobile communication networks, the main consumers of edge services have evolved from static to mobile devices, e.g., Unmanned Aerial Vehicles (UAVs). In this context, MS migration has gathered momentum as the key technology to ensure continuous proximity of latency-sensitive and bandwidth-consuming MSs to mobile end users.

In this paper, we focus on {\em stateful} migration, which is used whenever keeping track of an MS state is essential to  service continuity. In fact, despite the current trend favoring the development of stateless MSs, stateful MSs are still extremely common due to the complexity in refactoring legacy monolithic applications~\cite{furda2018migrating}. Furthermore, according to service-oriented architecture patterns~\cite{erl2016soa}, some essential stateful utility services will always be required, even if stateless service implementation will become dominant.

{\bf \em Existing Issues.} Although migration represents a powerful tool to ensure MSs proximity to the end users as they move, in practice, some service disruption during a migration process is unavoidable and must be accounted for. This is because: (i) stateful container migration techniques require freezing the MS state, and (ii) the network connection between the server running the MS and the mobile end users has to be migrated, along with the containerized MS.
Further, as discussed in detail in Sec.~\ref{work05:sec:related_work}, no existing migration framework enables an effective and efficient implementation of the stateful MS migration process at edge scale. 
We fill this gap by proposing a Migration Orchestration framework for microServices at the Edge (MOSE), designed to attain stateful migration of latency-critical edge MSs.

{\bf \em Technical Challenges.} Implementing an effective and efficient migration framework at the edge is challenging since:
\begin{itemize}
    \item[\emph{(i)}] the service disruption duration due to the migration process depends on several factors, such as coding optimization, communication protocols, and limited nodes' computational capabilities, thus its minimization is not trivial; 
    \item[\emph{(ii)}] upon statefully migrating an MS, the established connection thereof with the mobile end user needs to be preserved; 
    \item[\emph{(iii)}] the stateful migration process needs to be independent of the specific MS and of the underlying edge technology; 
    \item[\emph{(iv)}] to effectively actuate and configure the migration process, it has to be envisioned an orchestrator that can collect and aggregate all relevant metrics. 
\end{itemize}
To the best of our knowledge, no other work has jointly tackled all these four aspects at the same time.

{\bf \em Summary of Novel Contributions.}
We present MOSE, which implements stateful MS migration at the edge and orchestrates the migration process to ensure minimal impact on the user's QoE. Specifically, MOSE (i) enables the implementation of stateful migration of a generic MS at edge scale while (ii) preserving its connection with the mobile end user.
Importantly, (iii) its agnostic design enables a seamless and efficient integration on top of already existing edge platforms, (iv) it features several optimizations of the migration workflow, thus minimizing the experienced service disruption duration, and (v) leveraging the PAM model~\cite{calagna2024design}, it effectively configures the migration process, so that the target migration KPIs and the vertical's objectives are met, even in the presence of strict reliability and robustness requirements.

We validate MOSE using realistic MSs and, by experimentally comparing MOSE to other solutions in prior art, we show how it greatly reduces service disruption thanks to an effective migration orchestration. Moreover, to demonstrate the benefits of our solution in real-world scenarios, we consider two practical use cases featuring, respectively, a UAV autopilot MS and a Machine Learning (ML)-based multi-object tracking MS. We demonstrate how MOSE can be  exploited to  attain the MS state preservation upon migration and  effectively configure the migration process to guarantee a satisfying QoE, while greatly outperforming the state-of-the-art in terms of both time performance and resource utilization.

{\bf \em Paper organization.} Before introducing MOSE in   Sec.~\ref{work05:sec:framework}, we discuss some related work while highlighting the novelty of our study in Sec.~\ref{work05:sec:related_work}.  Sec.~\ref{work05:sec:overview}  provides some preliminaries to introduce stateful migration and our previous contributions, while  Sec.~\ref{work05:sec:exp_validation} and Sec.~\ref{work05:sec:mose_exploitation} validate and show the benefits of  MOSE.  Finally,  we draw our conclusions in Sec.~\ref{work05:sec:conclusions}.

\section{Related Work}\label{work05:sec:related_work}

There exists a considerable amount of literature regarding MS migration. Specifically, two fundamental migration techniques have been consolidated: stateless and stateful migration. The former is well-established and supported by most off-the-shelf container orchestration platforms, such as Proxmox~\cite{proxmox} and Kubernetes~\cite{kubernetes}. For instance, \cite{kaur2023live} migrates 5G core network functions across Kubernetes clusters by simply stopping and restarting containers, thus neglecting the containers' internal state.
Our work, instead, focuses on stateful migration of containers and migration of service-user connections, addressing the challenge of keeping track of an MS state, and the connection thereof with the end users, to guarantee service continuity.

Most of prior art related to service migration highlights the need for effective migration techniques that minimize service disruption in practical scenarios, e.g., \cite{dupont2017edge,liu2018migration,labriji2021mobility}, tackling, respectively, IoT tasks offloading, video streaming MSs, and vehicular mobility.
Focusing on stateful container migration technique, relevant examples of works include~\cite{terneborg2021application}, which gives an overview of current container migration techniques along with their fundamental metrics, and~\cite{berg2019evaluating, sindi2019using,puliafito2020impact,htet2021implementation}, which present promising applications of stateful migration by using CRIU. 
However, in spite of quite a large body of work in the field, few studies have addressed the implementation of a migration framework at the network edge. Among these, \cite{abdullaziz2019enabling} proposes a solution for mobile service continuity in edge-enabled WiFi networks and addresses service disruption minimization using a real-time operating system. \cite{ramanathan2021live}, instead, focuses on the components of the mobile core network and demonstrates that container PreCopy outperforms other migration strategies and virtualization technologies.
A proof-of-concept orchestration architecture is introduced in~\cite{muller2022architecture}, for improving fault-tolerance by leveraging container migration. Further, \cite{aleyadeh2022optimal, panek2023relocator} formulate optimization problems that aim, respectively, to achieve minimal downtime for fault recovery and to meet QoS requirements while relocating edge applications. \cite{Ray2024learning,Afrasiabi2023RLopt} demonstrate the effectiveness of reinforcement learning-based solutions to track user mobility and proactively migrate containers while jointly minimizing application latency and migration cost.

As for connection migration, many existing studies, e.g., \cite{bao2017follow,bellavista2019differentiated,an2019seamless} have tackled re-connection after a container migration. From a practical perspective, such an approach implies a customization of the client application source code to support the re-connection procedure. Only few works discuss solutions that enable connection migration in a completely transparent manner for the client and these are mostly based on dedicated protocols, network proxy, overlay network tunneling, and SDN. The studies in~\cite{qiu2017lxc,le2019experiences} propose MPTCP protocol as an effective solution, since it permits to define multiple sub-flows for the same connection. Nonetheless, MPTCP requires kernel customization, implying practical limitations in real-world scenarios. Similarly, \cite{raad2014achieving,puliafito2022server} respectively investigate and enhance LISP and QUIC protocol to effectively support connection migration, although yielding a custom protocol solution with limited generality. \cite{Goethals2024Warrens} proposes a decentralized, connectionless tunneling framework that enables communication between services deployed across public and private edge networks; however, it requires kernel customization to support ePBF technology. Other approaches \cite{kassahun2014pmip,benjaponpitak2020enabling,junior2022good} leverage either dedicated or cloud platform's network proxy to hold and redirect active connections upon service migration. Despite being quite simple and effective, the use of centralized proxies in unfit for latency-critical edge computing scenarios as it breaks the proximity principle with mobile end users.

At last, we mention that an initial version of this work has been presented in our conference paper~\cite{yu2024design}, sketching the design and implementation of MOSE. Here, we have significantly enhanced our contribution by (i) leveraging a more realistic testbed featuring a cloud computing architecture based on OpenStack~\cite{openstack}, (ii) thoroughly extending our experimental validation results, and (iii) proposing two practical exploitation scenarios to demonstrate the effectiveness of our solution compared to prior art, namely multi-object tracking and UAV autopilot MSs. 

{\bf Novelty.} Unlike existing works, MOSE enables an effective orchestration of the MS migration at the network edge while preserving the connection between the MS and the mobile end users. MOSE achieves this goal by leveraging and effectively combining (i) the PAM model to configure the migration process and (ii) our overlay network solution to enable connection migration, which we introduced in~\cite{calagna2024design}. MOSE builds upon  such solutions creating a novel, full-fledged algorithmic framework and technological solution that, differently from the existing alternatives, 
(i) {\em minimizes service disruption in a way that is application independent and requires neither dedicated protocol nor modifications to the kernel or application source code}, and (ii) {\em accurately configures and orchestrates the migration process of real-world MSs}, such as a UAV autopilot and a multi-object tracking MS, {\em while fulfilling both target KPIs and the vertical's objective}.

\section{Preliminaries: Used Technologies}\label{work05:sec:overview}
This section presents an overview of stateful container migration and the primary enabling tools to implement it. Further, it introduces COAT, our network solution to enable connection migration, and PAM, our model for the fundamental KPIs of stateful migrations~\cite{calagna2024design}. 

{\bf Stateful Container Migration.}
We focus on stateful migration, which enables the relocation across edge hosts of MSs containers whose internal state must be migrated to ensure service continuity. In other words, the migrated container can seamlessly restore its previous working state, thus guaranteeing minimal impact on the Quality of Experience (QoE) of the final users. The fundamental off-the-shelf tools required to implement stateful container migration are CRIU and Podman, as detailed below.

{\em CRIU~\cite{criu-cr}.}
Checkpoint/Restore In Userspace (CRIU) is widely considered the key tool for stateful migration from a process layer perspective. It implements two major procedures: (i) the {\em checkpoint procedure} that freezes a running process, collects its internal state, and encapsulates it into an image, and (ii) the {\em restore procedure} that creates a new process and restores its state by using a previously acquired checkpoint image. The latter includes: (i) the CPU-context state, e.g., the processes tree structure and the associated registers, (ii) the network sockets, (iii) the memory content, and (iv) the open file descriptors.

{\em Podman~\cite{podman}.} It is an open-source tool designed to develop, manage, and run containers and pods according to the Open Container Initiative (OCI) standards. Among all off-the-shelf container engines, e.g., Docker and LXC, Podman is the one featuring the strongest integration with CRIU, by directly leveraging its APIs and, thus, effectively supporting container migration at the microservice layer.

\begin{figure}[tb]
  \centering
  \includegraphics[width=\linewidth]{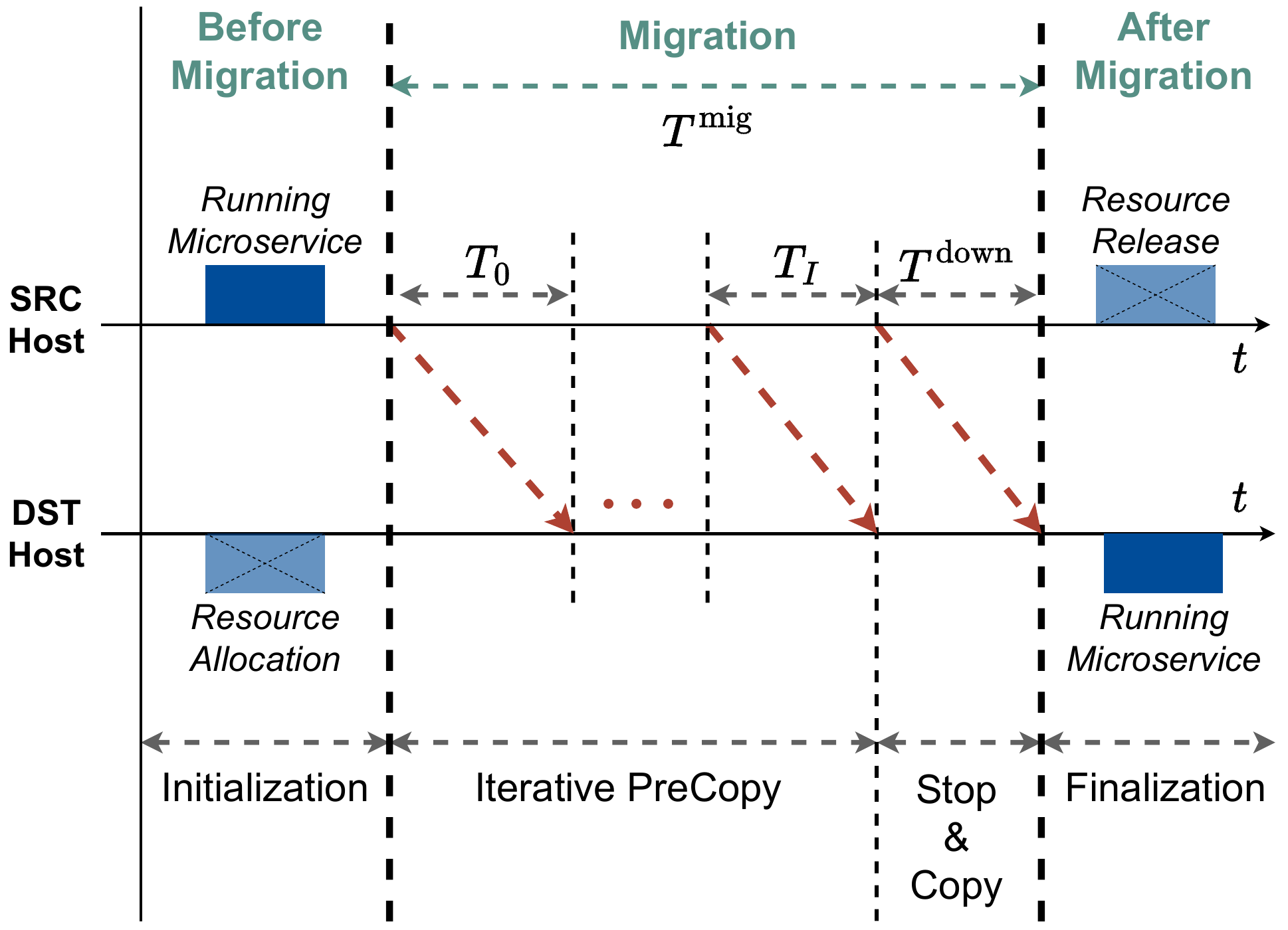}
  \caption{Stateful MS migration under the Iterative PreCopy strategy.}
  \label{work05:fig:it_precopy}
  \vspace{-5mm}
\end{figure}

Leveraging both CRIU and Podman, multiple stateful migration strategies can be defined. The most traditional and simplest one, i.e., {\em Cold Migration}, consists of the following steps: (1) creation of a checkpoint image at the source host, (2) transfer of such image from source to destination host, (3) restoration of the container at the destination host. Throughout these steps, the MS needs to be stopped at source host while it is yet to be restored at destination host, thus causing service disruption -- commonly referred to as ``downtime'' ($T^{\text{down}}$). To minimize the service disruption, the {\em Iterative PreCopy} strategy has been envisioned, which draws on the MS {\em dirty-page rate} concept, i.e., the number of memory pages a MS modifies per time unit. As depicted in Fig.~\ref{work05:fig:it_precopy}, this strategy consists of: (i) the iterative transfer of dirty pages to the destination host while the MS is still running at the source, (ii) a {\em Stop\&Copy} stage, during which the MS is stopped at the source host and the remaining dirty pages are transferred to the destination host where the MS will be eventually resumed. Hence, by minimizing the amount of data that needs to be transferred over the network, this approach allows trading a larger {\em total migration duration} ($T^{\text{mig}}$) for a shorter downtime, at the cost of an increased traffic burstiness.

\begin{figure}[tb]
  \centering
  \includegraphics[width=\linewidth]{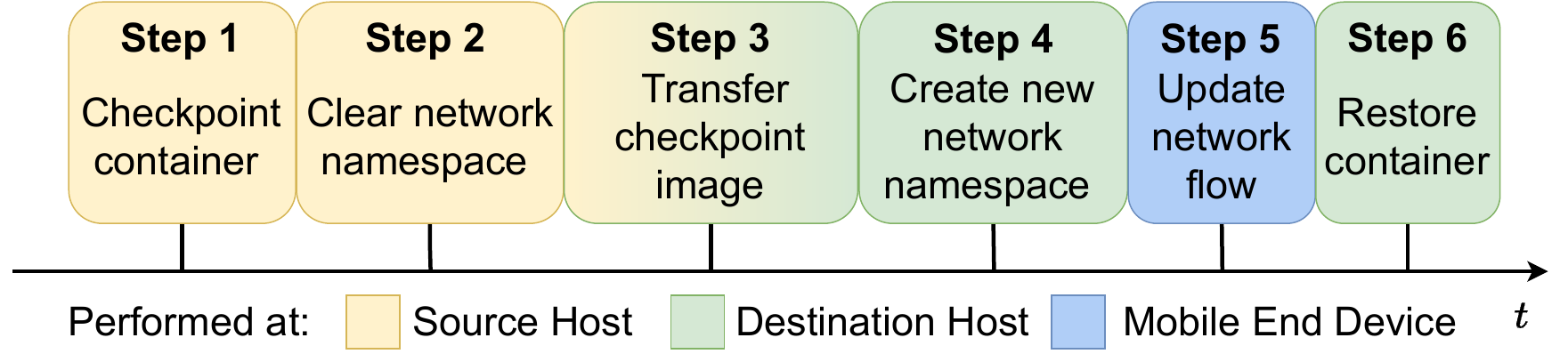}
  \caption{Enhanced Stop\&Copy stage for the stateful MS migration procedure integrating the COAT network solution~\cite{calagna2024design}.}
  \label{work05:fig:coat_procedure}
  \vspace{-3mm}
\end{figure}

{\bf COAT~\cite{calagna2024design}.} 
We proposed COAT, a  network solution that permits to preserve the connection established between an MS and the mobile end users during stateful migration. COAT is application independent, requires no dedicated protocol and no modifications to the kernel or application source code. 
To do so, it leverages Open vSwitch (OvS)~\cite{ovs} to define a proper logical  overlay network in which traffic flows can be dynamically managed.
Further, to effectively integrate COAT  with the traditional migration process, we introduced the COAT migration procedure, which includes an enhanced version of the Stop\&Copy stage of the stateful container migration process consisting of the steps illustrated in Fig.~\ref{work05:fig:coat_procedure}: (1) checkpoint the running container at the source host, thus collecting both the MS state and the established connection state; (2) clear the network namespace, thus preventing network configuration conflicts in the following steps; (3) transfer the checkpoint image from source to destination host; (4) re-create and configure the network namespace at the destination to match the original one, so that the later container restore procedure can successfully take place; (5) update the network flow of the connection by redirecting it towards the new network namespace; (6) restore the container from the checkpoint image. Now, the MS and its established connection can resume from their previous working state.

{\bf PAM Model~\cite{calagna2024design}.}
We developed PAM, an 
analytical model that effectively characterizes the fundamental stateful migration KPIs, i.e., migration duration and downtime, independently of the specific MS. PAM accounts for the processing time overhead introduced by the migration tool and its impact on the KPIs. In the following, we use PAM model to configure the migration process so as to fulfill the migration KPI targets and the vertical's objective, i.e., to either minimize migration downtime or resource usage in terms of bandwidth allocation and CPU consumption.

\section{The MOSE Framework}\label{work05:sec:framework}
This section presents our reference scenario and introduces our Migration Orchestration framework for microServices at the Edge (MOSE) along with its fundamental components.

\subsection{MOSE's overview and components}\label{work05:sub:solution}
\begin{figure}[tb]
  \centering
  \includegraphics[width=\linewidth]{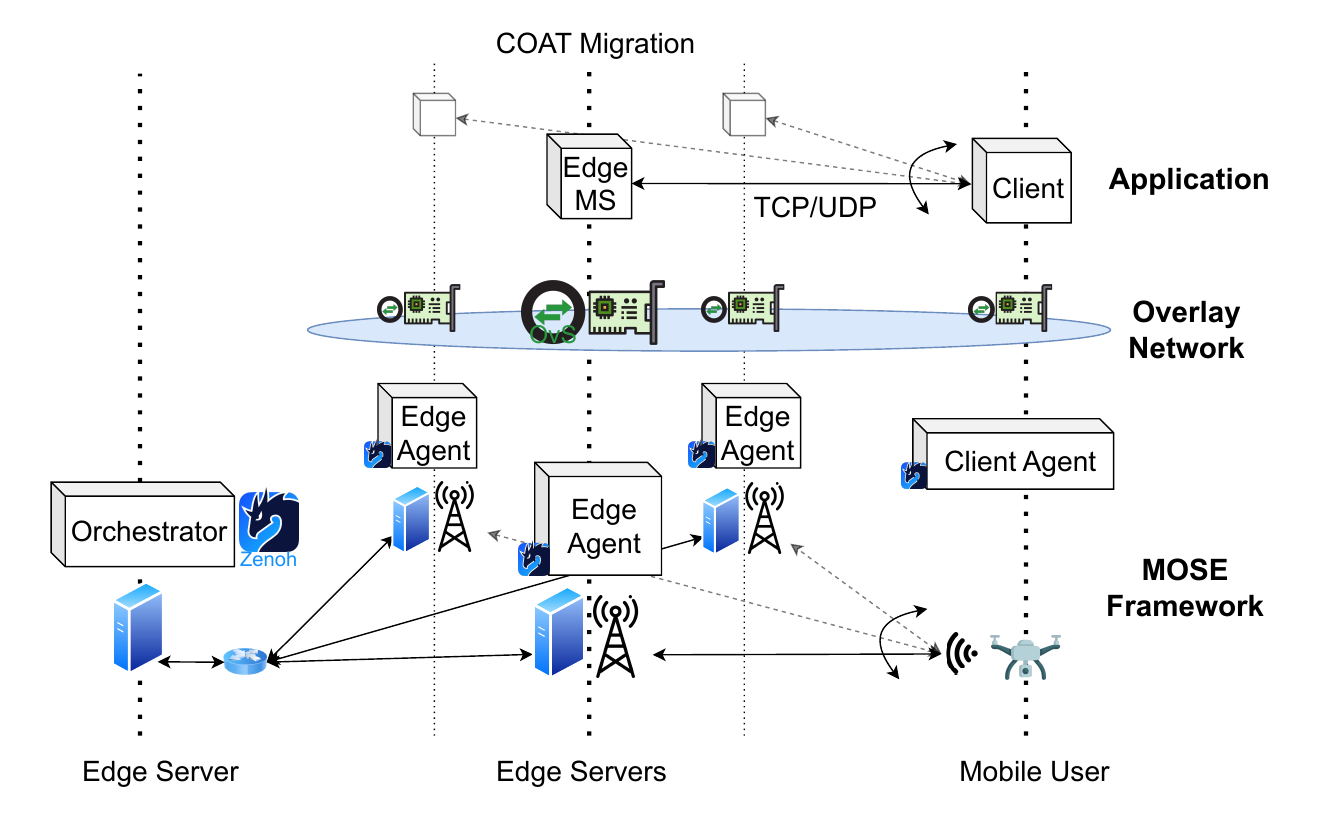}
  \caption{Reference system architecture integrating MOSE.}
  \label{work05:fig:mose_architecture}
  \vspace{-3mm}
\end{figure}

We start by introducing a simple, yet relevant, MS migration scenario, as depicted in Fig.~\ref{work05:fig:mose_architecture}. The example scenario includes a UAV as mobile device, connecting to different 5G base stations (gNBs) as it moves over the area of interest. Due to the UAV's limited computational resources, some of its critical functions (e.g., flight control with collision avoidance algorithm) must be offloaded at the edge through one or more MSs, hosted at an edge server with which the UAV connects using a given network protocol. To minimize the service latency, such MSs should be deployed on and, hence, migrated to, the edge server that is co-located with the gNB currently serving the UAV, as the latter changes its point of access. This fact, along with the need to maintain the internal state of the considered MS, makes stateful container migration the key technology to ensure the continuous fulfillment of the service requirements.

The MOSE framework is specifically designed to solve the above issue, i.e., to effectively configure and implement a stateful migration process while addressing all the relevant technical challenges:
(i) it minimizes service disruption, (ii) it preserves connectivity at the application level, (iii) it migrates an MS independently of the specific MS and the underlying edge technology, and (iv) it attains the migration KPI targets and vertical's objective. 

MOSE does so by leveraging three main components: 
a {\em migration orchestrator}, which properly configures the migration process;
a {\em migration agent} residing at each edge host and mobile device, which implements the MS migration;  
a {\em migration protocol} enabling the interaction between the migration orchestrator and the agents, as well as between the agents.
Fig.~\ref{work05:fig:mose_architecture} depicts the above scenario, mapping the MOSE components onto physical network entities. 
The MOSE orchestrator is deployed on an edge server which is responsible for monitoring and managing the overall edge system, while a MOSE edge agent is hosted at every other edge server, which may also be co-located with a gNB. The mobile device, which consumes the service deployed at one of the edge servers, hosts a MOSE client agent, which is responsible for  configuring the connection between the client  and the edge MS.

The migration protocol, which allows the agents to interact with each other as well as with the orchestrator, is based on Zenoh~\cite{zenoh}, a highly scalable~\cite{shih2022scalable} pub/sub/query protocol that provides extremely low latency and high throughput, substantially outperforming  popular communication protocols like MQTT, Kafka, NATS, and DDS~\cite{liang2023performance,zenoh-vs-nats,zhang2024comparison,baron2025zenoh}.
Given its simple mechanisms for Remote Procedure Calls (RPC) and low latency optimizations, Zenoh enables an effective and efficient message exchange between the migration entities. Also, its adaptive routing mechanism allows MOSE agents to dynamically and seamlessly join the framework at any time. As a result, our framework can adapt to diverse and evolving network topologies, including geo-distributed environments and scenarios with a large number of edge nodes.

As discussed in Sec.~\ref{work05:sec:overview}, our COAT migration process relies on an overlay network to preserve the network connection established between the MS and the mobile end user upon migration. We thus design the MOSE agent both to implement the COAT migration process {\em and} to configure the required overlay network using OvS.

In summary, from a high-level perspective, the application running on board of the mobile device (client application) interacts with an edge MS through a generic network connection. Leveraging MOSE, such MS, together with the connection established with the client application, can be migrated across different edge hosts in a transparent way with respect to both the client application and the MS itself. Further, leveraging the migration orchestrator, the migration procedure can be configured to satisfy (i) the migration KPI targets, thus guaranteeing minimum impact on the mobile device's QoE, and (ii) the vertical's objective (i.e., to either minimize the experienced service disruption or the resource consumption in terms of required network bandwidth and CPU usage). 

\begin{figure}[tb]
  \centering
  \includegraphics[width=\linewidth]{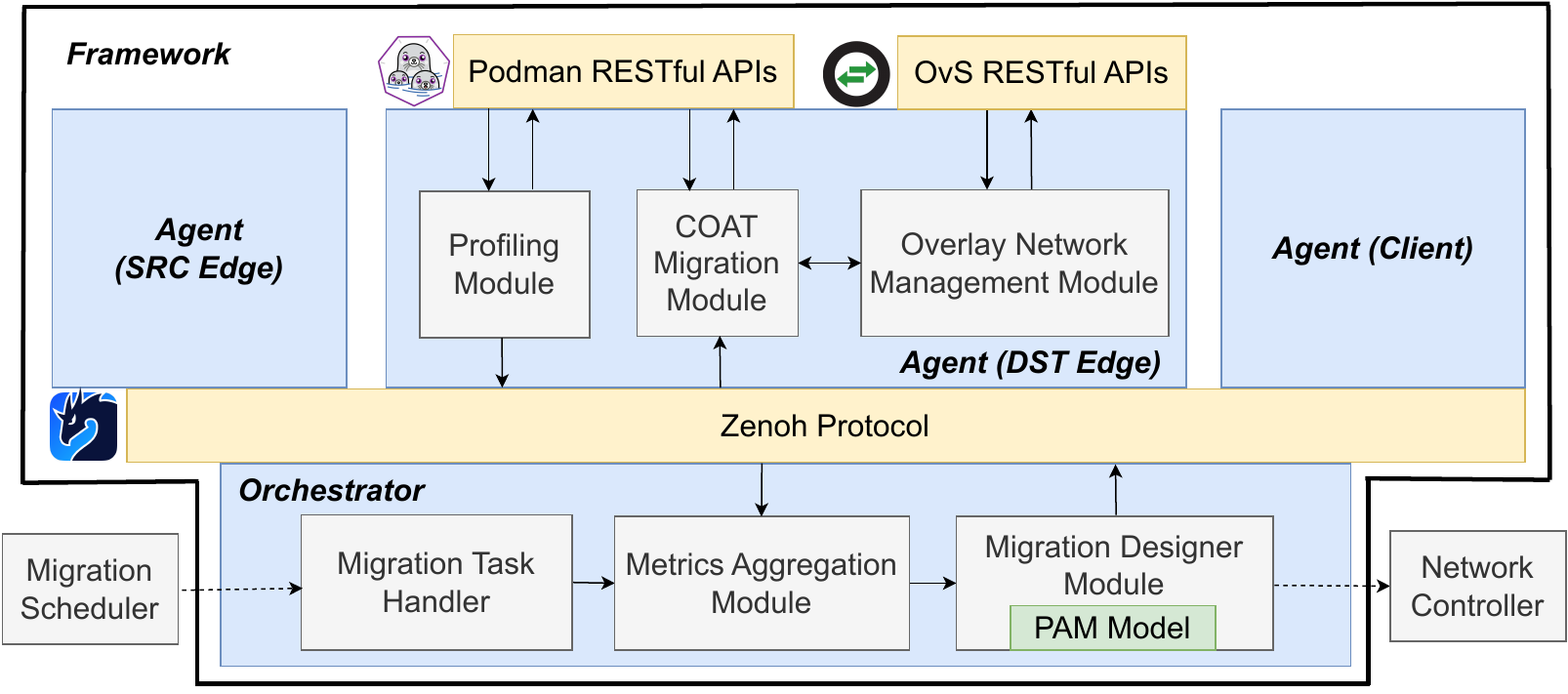}
  \caption{MOSE Agent and Orchestrator: modules and libraries.}
  \label{work05:fig:mose_framework}
  \vspace{-3mm}
\end{figure}

Below, we describe the fundamental characteristics of the MOSE components. Their structure and the employed libraries are also illustrated in Fig.~\ref{work05:fig:mose_framework}.

\begin{figure*}[!ht]
  \centering
  \includegraphics[width=\linewidth]{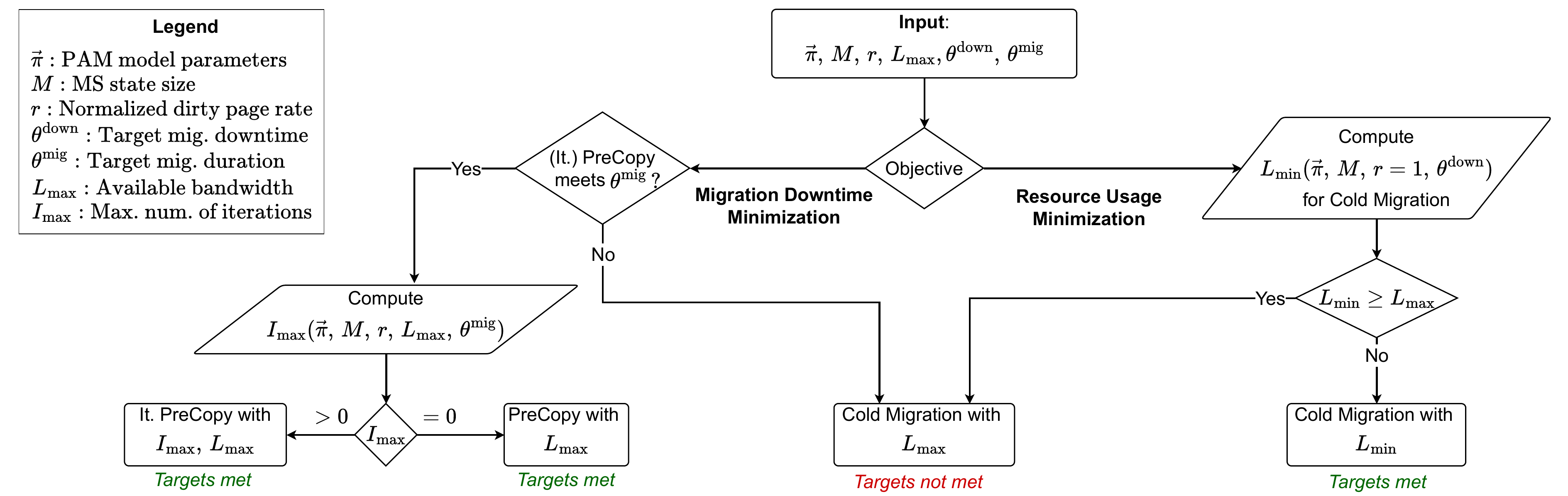}
  \caption{Configuration algorithm executed by the MOSE orchestrator.}
  \label{work05:fig:mose_algorithm}
  \vspace{-3mm}
\end{figure*}

\subsection{The MOSE agent}\label{work05:sub:mose_agent}
The MOSE agent runs at the mobile device and at each edge host; in the case of MS migration we refer to ``source agent" as the instance running in the source host and ``destination agent" as the one running in the destination host. It consists of three main modules handling, respectively, the COAT migration process, the overlay network, and the profiling mechanism.

The {\em COAT Migration module} is responsible for implementing the COAT migration process according to the configuration provided by the MOSE orchestrator. Then, the {\em Overlay Network Management module} is in charge of managing and configuring the overlay network, thus enabling our COAT migration strategy. Finally, the {\em Profiling module} allows the agent to characterize all the relevant aspects regarding migration, i.e., to estimate the available bandwidth on the links connecting edge hosts, measure the MS state size and dirty-page rate, and estimate the PAM model parameters. The value of available bandwidth can be obtained through an existing performance monitoring system at the edge, which is reasonable to assume is  in place in real-world deployments. Alternatively, it can be estimated by using Zenoh, which can foresee proper probes  from the source to the destination host.
Then, while the MS state size is attained by gathering the container memory usage through Podman, the MS dirty-page rate is derived leveraging the kernel's capabilities to keep track of memory changes~\cite{kernel-memory-track}. Specifically, we exploit CRIU to measure the number of memory pages $N_R$ that are modified per time unit $\Delta T$, yielding a dirty-page rate estimate $\hat{R}{=}N_R/\Delta T$. The normalized dirty-page rate $r$ is then computed as ${(\hat{R} {-} R_{\min})}/{(R_{\max} {-} R_{\min})}$, where $R_{\min}{=}1/\Delta T$ (measured in pages/s) and $R_{\max}{=} \lceil M/\sigma \rceil /\Delta T$, with $M$ and $\sigma$ denoting, respectively, the MS state size and the page memory size. Finally, PAM model parameters are estimated using DPRGen~\cite{calagna2024design}, a benchmark MS whose behavior in terms of memory allocation and dirty-page rate can be finely controlled. These parameters enable accurate modeling of the overall migration performance by capturing the impact of the computational capabilities allocated to each agent. As a result, the MOSE  orchestration algorithm is independent of specific resource configurations and is  inherently adaptive to diverse edge environments with heterogeneous capabilities. The resulting metrics, periodically collected, are sent to, and processed by, the MOSE orchestrator, running on an edge server that is responsible for monitoring and managing the overall edge system. Importantly, the computation and collection of these metrics can be performed in the background, with  negligible communication overhead and  impact on the overall migration performance.

The functionality of these agent modules strongly rely on the fundamental sets of RESTful APIs provided by Podman and OvS, both introduced in Sec.~\ref{work05:sec:overview}. While the former is used by Profiling and COAT Migration modules to interact with the Podman container engine, the latter is leveraged by the Overlay Network Management module to configure the overlay network. Finally, the message exchange between agent and orchestrator is regulated through the Zenoh protocol.

\subsection{The MOSE orchestrator}\label{work05:sub:mose_orchestrator}
The main modules composing the MOSE orchestrator are the migration task handler, the metrics aggregator, and the migration designer. 

The {\em Migration Task Handler} is responsible for processing the migration task whenever this is triggered by a migration scheduler, and it is able to handle   multiple migrations of heterogeneous MSs. The design and implementation of the migration scheduler are orthogonal to our solution and out of the scope of this work.
The handler assigns the migration task to MOSE by specifying:
(i) the ID of the container to be migrated, (ii) the IDs of the source and destination agents, and (iii) the target KPIs, along with the objective to be fulfilled, i.e., minimization of resource usage or minimization of migration downtime. Depending on this objective, the appropriate migration technique is selected and configured using our algorithm in the migration designer module described below.

The {\em Metrics Aggregation} module is instructed by the migration task handler about the set of information that has to be collected to fulfill the given task. 

The {\em Migration Designer} uses the aforementioned set of information and the PAM model to predict an upper bound on the migration KPIs and to properly set the migration parameters with the aim to fulfill the required level of QoE.
This module executes the orchestration algorithm, summarized in Fig.~\ref{work05:fig:mose_algorithm}, defining a configuration that specifies (i) which migration strategy to adopt (i.e., Cold Migration, PreCopy, or Iterative PreCopy), (ii) the network bandwidth that the network controller must allocate for the migration process, and, in case of Iterative PreCopy, (iii) the number of iterations to execute. To apply such configuration, the Designer module conveys it to the COAT migration module of both source and destination agents, using Zenoh protocol.

The algorithm input is given by: (i) the metrics and parameters retrieved through the profiling module from the source and destination agents, and (ii) the migration task as defined by the migration task handler, including the target KPIs and the driving vertical's objective.
When the objective is to minimize resource usage (see the right branch), the Cold Migration strategy is adopted, as it is the one that minimizes the amount of required bandwidth and CPU consumption. The algorithm then uses the PAM model to compute the minimum required bandwidth to meet the target on the downtime and checks whether it exceeds or not the estimated available one. When instead the goal is to minimize the migration downtime (see the left branch), the Iterative PreCopy strategy is enacted, as it minimizes the service disruption. The algorithm preliminary checks whether this strategy meets the target migration duration and, if so, it computes the maximum number of iterations to meet such target. Importantly, in the PAM model we consider the worst-case situation, i.e., the MS dirty-page is set to its maximum value, so that we obtain an upper bound to the migration and downtime duration. This allows neglecting the dependency on the specific iteration and, hence, a practical, yet accurate, safe way to configure the migration process.

To summarize, the algorithm outputs the migration configuration, consisting of (i) the migration strategy the agents should adopt, (ii) the minimum amount of network bandwidth that the network controller should reserve for the migration process, and, (iii) the number of PreCopy iterations to run, in case such strategy is selected.

\begin{figure}[tb!]
  \centering
  \includegraphics[width=\linewidth]{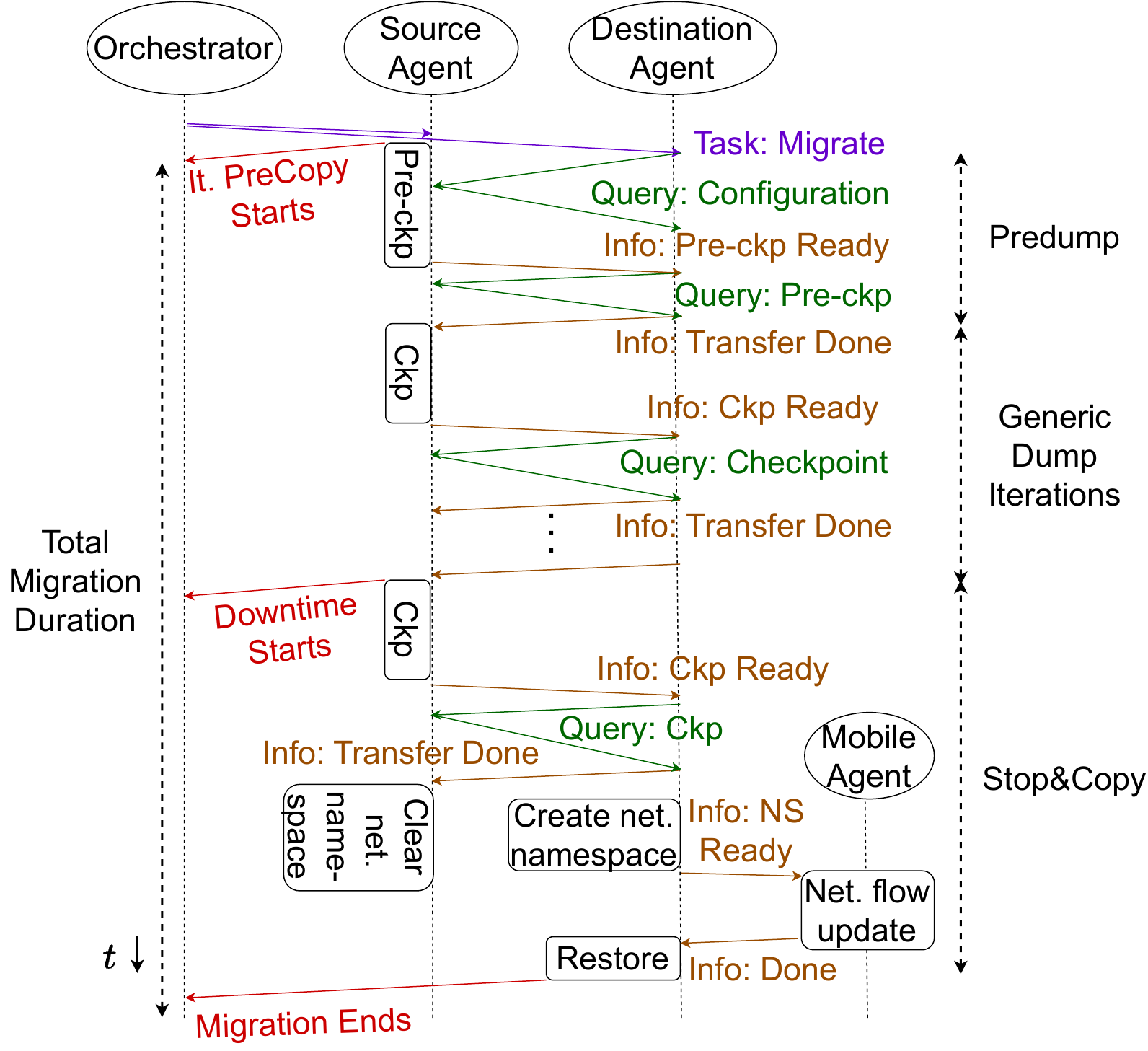}
  \caption{MOSE protocol under Iterative Precopy strategy and COAT Stop\&Copy.}
  \label{work05:fig:mose_mig_protocol}
  \vspace{-4mm}
\end{figure}

\subsection{The MOSE protocol}\label{work05:sub:mose_protocol}
The MOSE protocol is implemented via pub/sub/query mechanisms provided by Zenoh. It regulates: (i) the signaling of migration tasks, (ii) the actuation of the steps of the migration process, and (iii) the transfer of checkpoint images.

As an example, Fig.~\ref{work05:fig:mose_mig_protocol} illustrates the communication flow between the orchestrator and the agents when an Iterative PreCopy migration task is performed.
The migration task is published on a Zenoh topic and is accessed by the source and destination agents. First, the source agent informs the orchestrator that migration has begun. Then, while the source agent starts the pre-checkpoint procedure, the destination agent queries the required information to restore locally the MS, e.g., the MS's network namespace configuration setting. After the pre-checkpoint, the source agent asks the destination agent to gather the related pre-checkpoint image. Similarly, the source agent runs the checkpoint stage and notifies the destination host, which queries the corresponding image. After the checkpoint image is transferred, the MS's network namespace at the source host is cleared, while the namespace is recreated and configured, simultaneously, at the destination host. 

Upon being informed of the migration of the MS's network namespace, the mobile agent updates the client's network flow towards the agent, thus preserving the MS established connection with the client application. Finally, the destination agent restores the MS and notifies the orchestrator accordingly.

We remark that our communication solution enables an automated migration process with no need for additional remote control protocols such as SSH, thus minimizing the migration latency.

\section{Experimental Testbed and MOSE's Validation}\label{work05:sec:exp_validation}

Below, we first describe the testbed we developed and the configuration used for our experiments (Sec.~\ref{work05:sub:testbed}). Then, we evaluate the performance of our framework (Sec.~\ref{work05:sub:performance}) and show the effectiveness of our orchestration algorithm (Sec.~\ref{work05:sub:orchestration}).

\begin{table*}[!ht]
    \centering
    \caption{SockPerf MS migration performance comparison between SotA and MOSE framework for the COAT procedure steps in Fig.~\ref{work05:fig:coat_procedure}.}
\begin{tabular}{|l|c|c|c|c|c|c|}
    \hline
        \multirow{2}{*}{\textbf{\diagbox{Duration}{Scenario}}} & \textbf{\makecell{SotA: It. Precopy}} & \textbf{\makecell{SotA: Cold Mig.\ }}  & \multicolumn{2}{|c|}{\textbf{\makecell{MOSE-MD}}}  & \multicolumn{2}{|c|}{\textbf{\makecell{MOSE-MR}}}  \\ \cline{2-7}
         & 90\% C.I. [ms] & 90\% C.I. [ms] & 90\% C.I. [ms] & Decrease [\%] & 90\% C.I. [ms] & Decrease [\%] \\ \hline
        \textbf{S1: Checkpoint}                 & $1296 \pm 36$      & $1384 \pm 22$  & $503 \pm 17$   & 61     & $567 \pm 19$   & 59 \\
        \textbf{S2-4: Clear\&Create NS}           & -                     & -                 & $ 84 \pm 2$   & -         & $97 \pm 2$    & - \\ 
        \textbf{S3: Transfer}                   & $1057 \pm 11$      & $1267 \pm 6$   & $3.0 \pm 0.3$      & 99     & $93 \pm 1$    & 93 \\ 
        \textbf{S5: Flow Update}                & -                     & -                 & $4.0 \pm 0.1$      & -         & $4.0 \pm 0.1$     & - \\ 
        \textbf{S6: Restore}                    & $945 \pm 24$       & $934 \pm 10$   & $353 \pm 11$   & 63     & $353 \pm 12$   & 62 \\ \hline
        \textbf{Downtime, $T^{\text{down}}$}    & $3299 \pm 39$      & $3638 \pm 25$  & $975 \pm 21$   & 70     & $1323 \pm 37$  & 64 \\ 
        \textbf{Total, $T^{\text{mig}}$}        & $28487 \pm 192$    & $3638 \pm 25$  & $2986 \pm 46$  & 90     & $1323 \pm 37$  & 64 \\ \hline
    \end{tabular}
    \label{work05:tab:sockperf_mig_performance}
\end{table*}

\begin{table*}[!ht]
    \centering
    \caption{iPerf3 MS migration performance comparison between SotA and MOSE framework for the COAT procedure steps in Fig.~\ref{work05:fig:coat_procedure}.
    }
    \begin{tabular}{|l|c|c|c|c|c|c|}
    \hline
         \multirow{2}{*}{\textbf{\diagbox{Duration}{Scenario}}} & \textbf{\makecell{SotA: It. Precopy}} & \textbf{\makecell{SotA: Cold Mig.}}  & \multicolumn{2}{|c|}{\textbf{\makecell{MOSE-MD}}}  & \multicolumn{2}{|c|}{\textbf{\makecell{MOSE-MR}}}  \\ \cline{2-7}
         & 90\% C.I. [ms] & 90\% C.I. [ms] & 90\% C.I. [ms] & Decrease [\%] & 90\% C.I. [ms] & Decrease [\%] \\ \hline
        \textbf{S1: Checkpoint}                 & $1362 \pm 31$  & $1418 \pm 23$      & $345 \pm 11$   & 75     & $374 \pm 15$   & 74 \\ 
        \textbf{S2-4: Clear\&Create NS}           & -                 & -                     & $93 \pm 1$     & -         & $92 \pm 2$    & - \\ 
        \textbf{S3: Transfer}                   & $1045 \pm 10$   & $1190 \pm 4$       & $4.0 \pm 0.2$      & 99     & $8.0 \pm 0.3$     & 99 \\ 
        \textbf{S4: Flow Update}                & -                 & -                     & $3.0 \pm 0.1$      & -         & $4.0 \pm 0.1$     & - \\ 
        \textbf{S5: Restore}                    & $1029 \pm 128$ & $902 \pm 8$        & $331 \pm 10$    & 68     & $345 \pm 24$   & 62 \\ \hline
        \textbf{Downtime, $T^{\text{down}}$}    & $3437 \pm 134$ & $3562 \pm 26$      & $790 \pm 14$   & 77     & $1043 \pm 38$   & 71 \\ 
        \textbf{Total, $T^{\text{mig}}$}        & $32100 \pm 157$& $3562 \pm 26$      & $1837 \pm 25$  & 94     & $1043 \pm 38$   & 71 \\ \hline
    \end{tabular}
    \label{work05:tab:iperf_mig_performance}
    \vspace{-3mm}
\end{table*}

\begin{figure}[tb]
    \centering
    \includegraphics[width=0.75\linewidth]{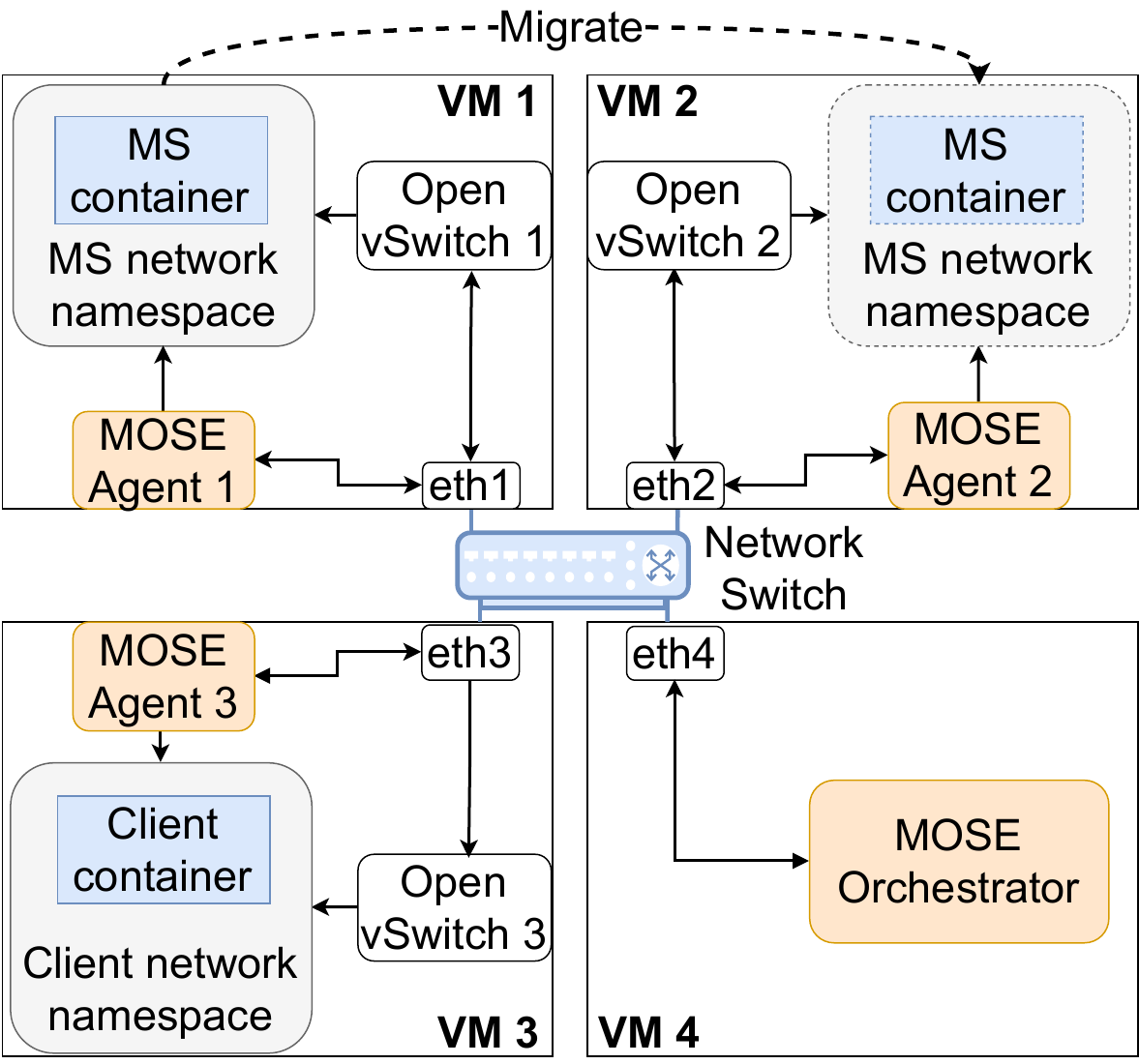}
    \caption{Testbed setup for the MOSE framework.}
    \label{work05:fig:mose_testbed}
    \vspace{-3mm}
\end{figure}

\subsection{Testbed setup}\label{work05:sub:testbed}
The testbed we developed to validate MOSE  comprises four identical virtual machines (VMs), each with   4\,vCPUs and 16\,GB of RAM and hosted on a cloud computing architecture based on OpenStack~\cite{openstack} and featuring Intel Xeon Skylake CPU. As depicted in Fig.~\ref{work05:fig:mose_testbed}, VM1 and VM2 represent two edge servers, acting, respectively, as the source and destination of the migration process. VM3 hosts the client container, thus acting as an end device that interacts with the edge servers, while VM4 acts as the edge server in charge of monitoring and managing the overall edge system.

To build our framework, we deploy three independent instances of the MOSE agent, respectfully on VM1, VM2, and VM3, and one MOSE orchestrator on VM4. Whenever a MOSE agent is initiated, it first measures the properties of the host machine through the profiling module and registers to the MOSE orchestrator using Zenoh protocol. Unlike the testbed in our previous study~\cite{yu2023tcp}, which depends on scripts and SSH tunnels to deliver the migration commands, thus being inefficient and complex to configure, the MOSE framework leverages the discovery mechanism implemented by Zenoh (see Sec.~\ref{work05:sub:solution}) to automatically create always-established connections among agents and orchestrator. Such enhancement brought by the MOSE framework significantly simplifies the testbed deployment and improves the scalability of the solution.

We conducted two independent sets of experiments employing SockPerf~\cite{sockperf} and iPerf3~\cite{iperf} as MSs to migrate. SockPerf is a network latency benchmarking tool to measure the communication latency in {request-reply} connectivity tests on the client side, while iPerf3 is a popular tool for active measurements of the achievable bandwidth on IP networks. Both MSs are meaningful examples of stateful MSs since their internal state comprises multiple counters and accumulators, which need to be preserved to attain consistent benchmarking results upon migration.
Further, they resemble real-world MSs with an established TCP connection and their features as network measurement tools allow us to effectively assess the impact of the migration process on the user QoE. For both SockPerf and iPerf3 experiments, we deploy their server-side at source host (VM1), and their client-side on the mobile device (VM3). Leveraging the profiling module (see Sec.~\ref{work05:sub:mose_agent}), we characterize the MSs in terms of state size and dirty-page rate under default settings. Specifically, we measured the state size of the SockPerf and iPerf3 MSs containers respectfully equal to 10\,MB and 0.5\,MB. Furthermore, we observed a dirty-page rate for both MSs containers ranging from 3 to 5 pages/s, which, given a default page-size of 4096\,B, is a relatively low value compared to their state size.

According to the mobility scenario shown in Fig.~\ref{work05:fig:mose_architecture}, we first consider a mobile device (VM3) moving away from the gNB co-located with the source edge host (VM1) and approaching the one co-located with the destination host (VM2); the increased latency triggers the migration process. Such radio link does not impact migration performance but we preserve the established connection  between the MS and the mobile device application. Further, we consider a network slice dedicated to the migration process to ensure a high-data rate network connection between the source and destination hosts.

\begin{figure*}[htp]
    \centering
    \subfloat[]{\label{work05:fig:vary_duration_iter}{\includegraphics[width=0.27\textwidth]{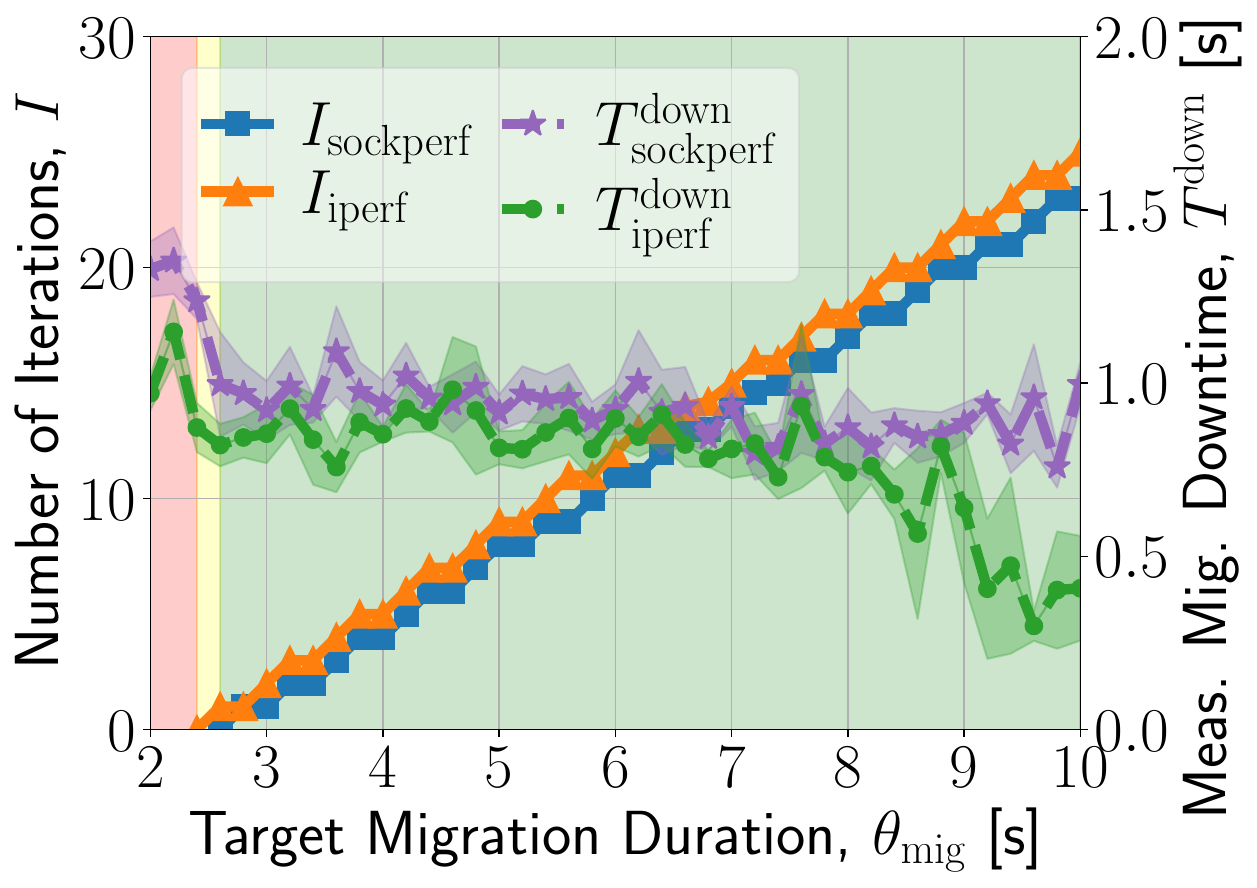}}}
    \subfloat[]{\label{work05:fig:vary_duration_result}{\includegraphics[width=0.24\textwidth]{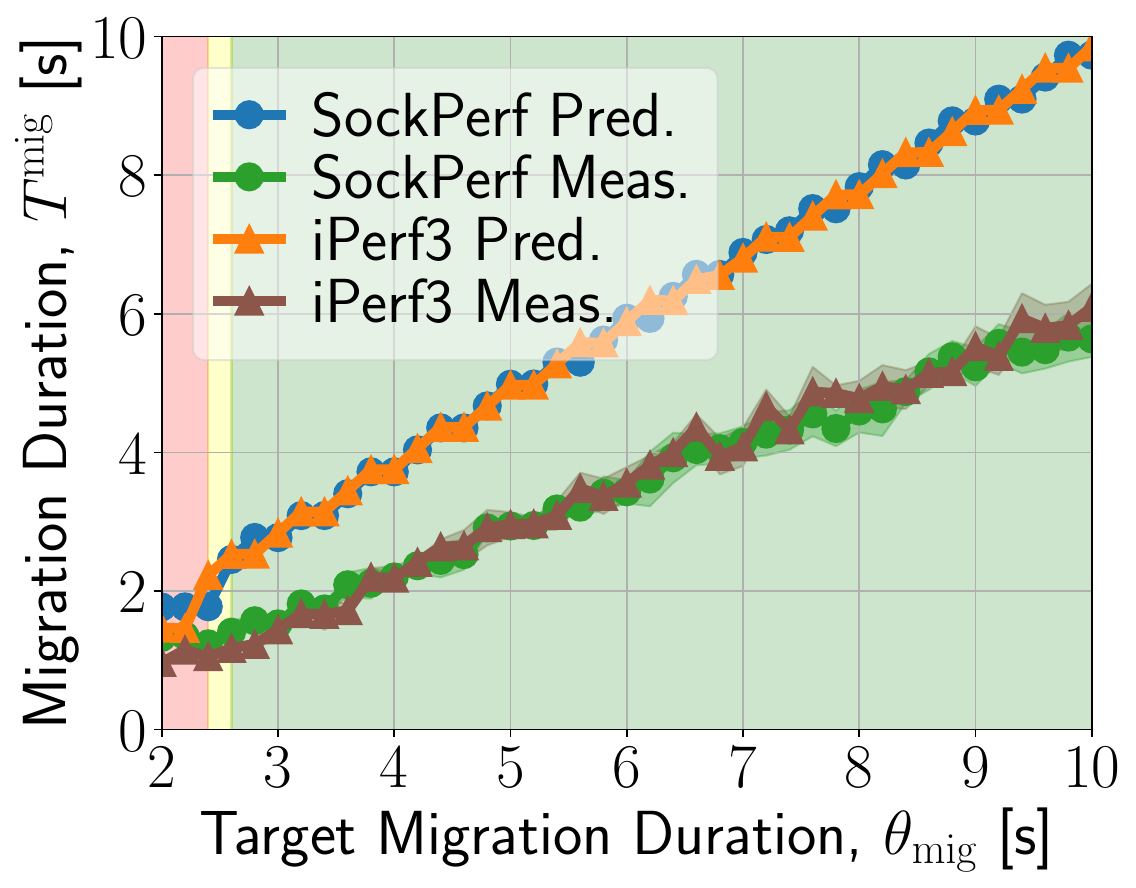}}}
    \subfloat[]{\label{work05:fig:vary_downtime_bandwidth}{\includegraphics[width=0.25\textwidth]{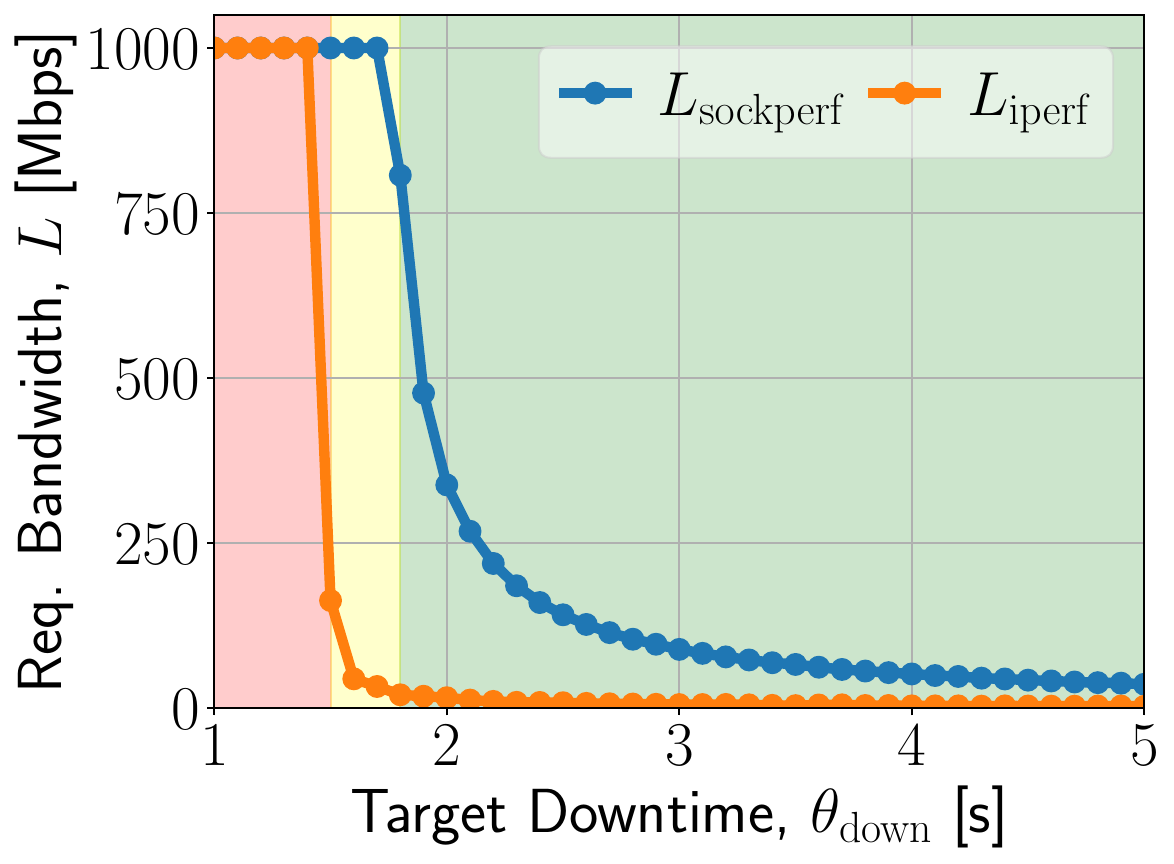}}}
    \subfloat[]{\label{work05:fig:vary_downtime_result}{\includegraphics[width=0.235\textwidth]{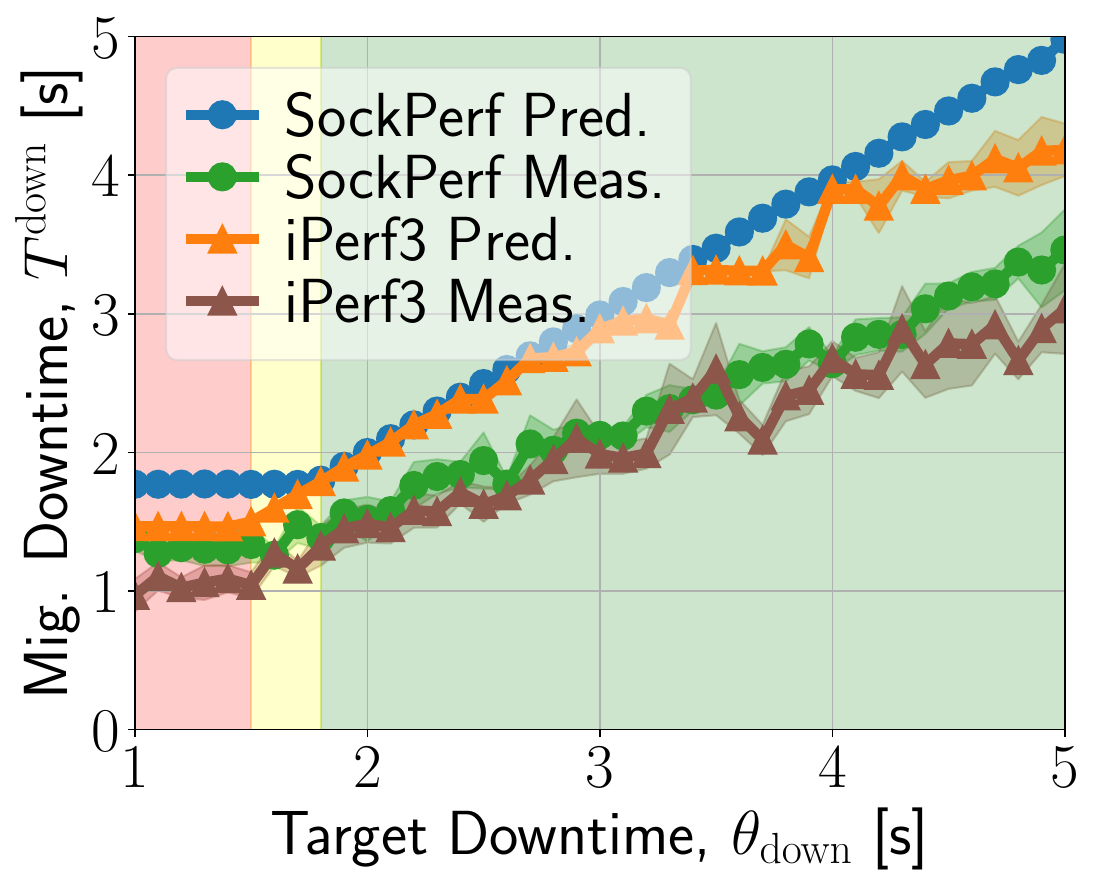}}}
    \caption{MOSE-MD (a)-(b) and MOSE-MR (c)-(d) performance. (a) Number of iterations for varying target migration duration and measured downtime; (b) Predicted and measured migration duration for varying target values; (c) Required network bandwidth for varying target downtime; (d) Predicted and measured downtime for varying target values.}
    \label{work05:fig:obj_results}
    \vspace{-3mm}
\end{figure*}

\subsection{MOSE migration performance}\label{work05:sub:performance}
Leveraging our testbed, we experimentally characterize each step of the migration process, for both SockPerf and iPerf3, accounting for different strategies.
We consider the state-of-the-art (SotA) migration strategies, which rely on scripts and SSH tunnels to control the migration procedure, alongside two variants of MOSE framework, depending on the driving vertical's objective that is considered: MOSE-MD to Minimize Downtime (MD) and MOSE-RM to Minimize Resource usage (MR). 
All results have been obtained by considering the 90\% confidence interval over 100 repetitions, and providing the average time improvement under MOSE with respect to the SotA. Specifically, we compare Iterative PreCopy and Cold migration strategies as defined in prior art with MOSE-MD and MOSE-MR, respectively. Further, for the MOSE experiments, we respectively set the target migration duration at 5\,s and the target migration downtime at 2\,s, yielding a configured network bandwidth equal to the maximum available one, i.e., 1\,Gbps, and a number of iterations $I{=}8$ for SockPerf and $I\mathord{=}9$ for iPerf3.

The results are shown in Table~\ref{work05:tab:sockperf_mig_performance} and Table~\ref{work05:tab:iperf_mig_performance}. They report the operational duration of each container migration step (as shown in Fig.~\ref{work05:fig:coat_procedure}), along with the fundamental migration KPIs, i.e., the migration downtime $T^{\text{down}}$ and the total migration duration $T^{\text{mig}}$. We remark that, compared to the results in Fig.~\ref{work05:fig:coat_procedure}, the duration of the migration steps S2 and S4 are reported in a combined way, as such steps are executed in parallel. Further, these steps, along with S5, refer to our COAT network solution to achieve connection migration~\cite{calagna2024design}. Instead, most prior art leverages proxy-based solutions, which do not feature any of these steps related to namespace and overlay network management. Importantly,  such additional COAT steps have negligible impact on the final KPIs, thus demonstrating the effectiveness of our solution, which, we recall, enables connection migration in a way that is application independent and requires neither dedicated protocol nor modifications to the kernel or application source code.

Looking at both tables and by focusing on the duration of the migration steps (S1 to S6), it can be noticed that the most significant time improvement brought by MOSE relatively to the SotA is on S3 (Transfer step), corresponding to a reduction of approximately 100\% for both SockPerf and iPerf3 MSs. Such an improvement is due to our efficient approach to signaling, according to which MOSE completely avoids the time overhead due to SSH, which is quite substantial, especially for low values of state size. In addition, comparing the results in the two tables, most of the values in Table~\ref{work05:tab:iperf_mig_performance} are smaller than those in Table~\ref{work05:tab:sockperf_mig_performance}, which is mainly due to the smaller state size characterizing iPerf3 container.

Consider now the measurements for the single MSs.
From Table~\ref{work05:tab:sockperf_mig_performance}, it can be noticed that S3 (Transfer step) highlights a significant time difference between the two scenarios. In fact, when minimizing resource usage, MOSE-MR employs the Cold migration strategy, which performs a checkpoint of the entire container at once and computes the minimum amount of network bandwidth that allows to transfer such checkpoint image while fulfilling the target downtime. Conversely, when minimizing downtime, MOSE-MD  configures the agents to use the Iterative PreCopy migration strategy. Therefore, during the final Stop\&Copy step, only dirty memory pages and the CPU-context are included in the checkpoint image, significantly reducing its size compared to the whole MS state size. Additionally, besides using the more efficient Zenoh protocol in place of SSH tunnels, agents utilize the maximum available network bandwidth for transferring checkpoint images, resulting in a minimal transfer duration. At last, by looking at the operation time of the other migration steps (i.e., S1, and S6), the MOSE framework achieves a reduction by up to  63\% compared to the SotA.

On the other hand, observing the operation time for S3 in Table~\ref{work05:tab:iperf_mig_performance}, no significant time difference between MOSE-MR and MOSE-MD can be observed. This is due to the fact that iPerf3 MS features both significantly small state size and dirty-page rate, yielding a comparable state transfer duration regardless of the vertical's objective. Again,  MOSE reduces the operation time of the other migration steps (i.e., S1, and S6) by up to  75\%.

The last two rows of Tables~\ref{work05:tab:sockperf_mig_performance} and~\ref{work05:tab:iperf_mig_performance} present the values of the fundamental migration KPIs. Notably,  MOSE  achieves a significant improvement in migration downtime, up to approximately 71\% and 77\% for, respectively, SockPerf and iPerf3 MSs. {\em Such significant reduction on the KPIs and on each migration step, allows passing from second to sub-second operations, thus making stateful container migration suitable for time-critical MSs.}
We recall that this improvement is due to two key features of MOSE: (i) its efficient approach to signaling, which completely avoid the time overhead due to SSH; (ii) the direct interaction of the agents with the Podman and OvS APIs, resulting in a more efficient command execution.

\subsection{MOSE orchestration strategy}\label{work05:sub:orchestration}

We now demonstrate the effectiveness of our MOSE orchestration algorithm under the two vertical's objectives adopted in MOSE. To do so, for MOSE-MD  we vary the target migration duration, as  in Figures~\ref{work05:fig:vary_duration_iter}-\ref{work05:fig:vary_duration_result}, and for MOSE-MR the target downtime, as in Figures~\ref{work05:fig:vary_downtime_bandwidth}-\ref{work05:fig:vary_downtime_result}. The results show the configuration output by the orchestrator, along with the values obtained for the fundamental migration KPIs.
Importantly, depending on the target values, we identify three main regions where the given target can be met (i) for neither SockPerf nor iPerf3 (red), (ii) for iPerf3 only (yellow), and (iii) for both MSs (green). 

{\bf Migration downtime minimization.} To attain this goal, MOSE-MD configures the migration process so as to leverage the maximum available bandwidth for the link connecting source and destination edge hosts, and it computes the number of PreCopy iterations that allows meeting the target migration duration. We validate our orchestration algorithm by varying the target migration duration from 2\,s to 10\,s, and observe, for both SockPerf and iPerf3, the resulting migration configuration, along with the KPIs values that are actually experienced. Specifically, Fig.~\ref{work05:fig:vary_duration_iter} shows the number of PreCopy iteration $I$ and the corresponding migration downtime $T^{\text{down}}$ as functions of the target migration duration $\theta_{\text{mig}}$.
When $\theta_{\text{mig}}$ cannot be met (red region), the MOSE orchestrator configures the migration process according to the Cold migration strategy, thus resulting in a relatively high migration downtime. Instead, when the migration duration target can be met (green region), the MOSE orchestrator selects the Iterative PreCopy migration strategy and increases the number of iterations $I$ consistently with the target. Due to the iPerf3 smaller state size, the Iterative PreCopy strategy is feasible for smaller values of $\theta_{\text{mig}}$ relatively to SockPerf, resulting in a performance in the yellow region. 

Further, Fig.~\ref{work05:fig:vary_duration_result} depicts the total migration duration for varying values of $\theta_{\text{mig}}$.
Specifically, it shows the upper bound that is predicted using the PAM model and the actual migration duration, both of which exhibit a behavior that is consistent with the configured number of iterations. Notably, the measured migration duration is always shorter than the prediction thereof, thus validating the capability of PAM of providing an upper bound for such KPI. However, due to the accumulation of the error, the gap between the predicted and actual migration durations becomes more evident as the number of PreCopy iterations increases.

\begin{figure}[tb]
  \centering
  \includegraphics[width=\linewidth]{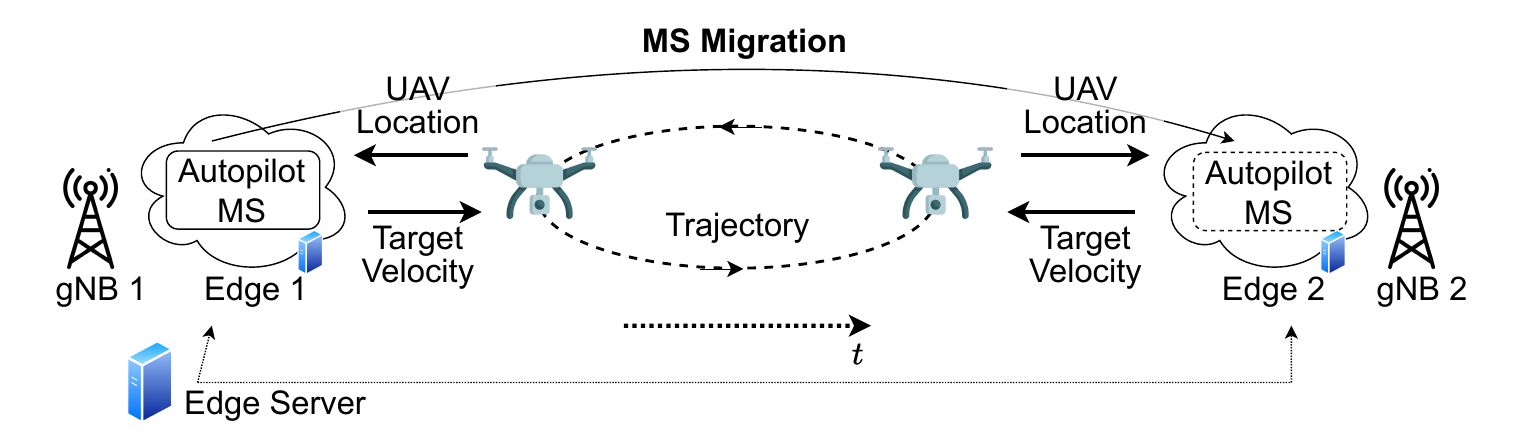}
  \caption{UAV autopilot MS migration reference scenario.}
  \label{work05:fig:autopilot_scenario}
  \vspace{-5mm}
\end{figure}

\textbf{Resource usage minimization.} To achieve this goal, MOSE-MR selects Cold Migration and computes the minimum bandwidth, $L$, required on the link connecting the source and destination edge hosts that fulfills the target downtime. To validate our orchestration algorithm, we now vary the target downtime from 1\,s to 5\,s, and, again, we record the migration configuration and the KPIs values for both SockPerf and iPerf3.

Fig.~\ref{work05:fig:vary_downtime_bandwidth} shows $L$ as a function of the target downtime $\theta_{\text{down}}$. When $\theta_{\text{down}}$ cannot be met (red region), MOSE configures the migration to use the maximum available network bandwidth, i.e., 1\,Gbps. Otherwise (green region), it computes through the PAM model the minimum amount of network bandwidth between source and destination hosts that meets the corresponding target. In fact, as $\theta_{\text{down}}$ increases, the value of $L$ decreases. Since the iPerf3 container has a smaller state size than the SockPerf container, the reduction of the required bandwidth $L$ is feasible for smaller values of $\theta_{\text{down}}$ (yellow region).
Furthermore, Fig.~\ref{work05:fig:vary_downtime_result} presents the upper bound that is computed through the PAM model and the value that is actually experienced, as $\theta_{\text{down}}$ varies. Both values exhibit a negative correlation with the value of $L$. Again, the measured value of downtime is always smaller than the predicted one, which underlines how the PAM model provides an accurate upper bound on such KPI.

{\bf Summary}. Our results highlight that MOSE: (i)   greatly improves  the migration performance relatively to the state of the art (in both its considered variants, i.e., MOSE-MD and MOSE-MR); (ii) reduces the duration of each migration step from seconds to sub-seconds, thus making migration feasible for time-critical applications; (iii)  effectively configures  the migration process to meet the KPI target values and accounting for different vertical's objectives.

\begin{figure}[tb]
  \centering
  \subfloat[MOSE-MD]{\label{work05:fig:uav_min_down}{\includegraphics[width=0.7\linewidth]{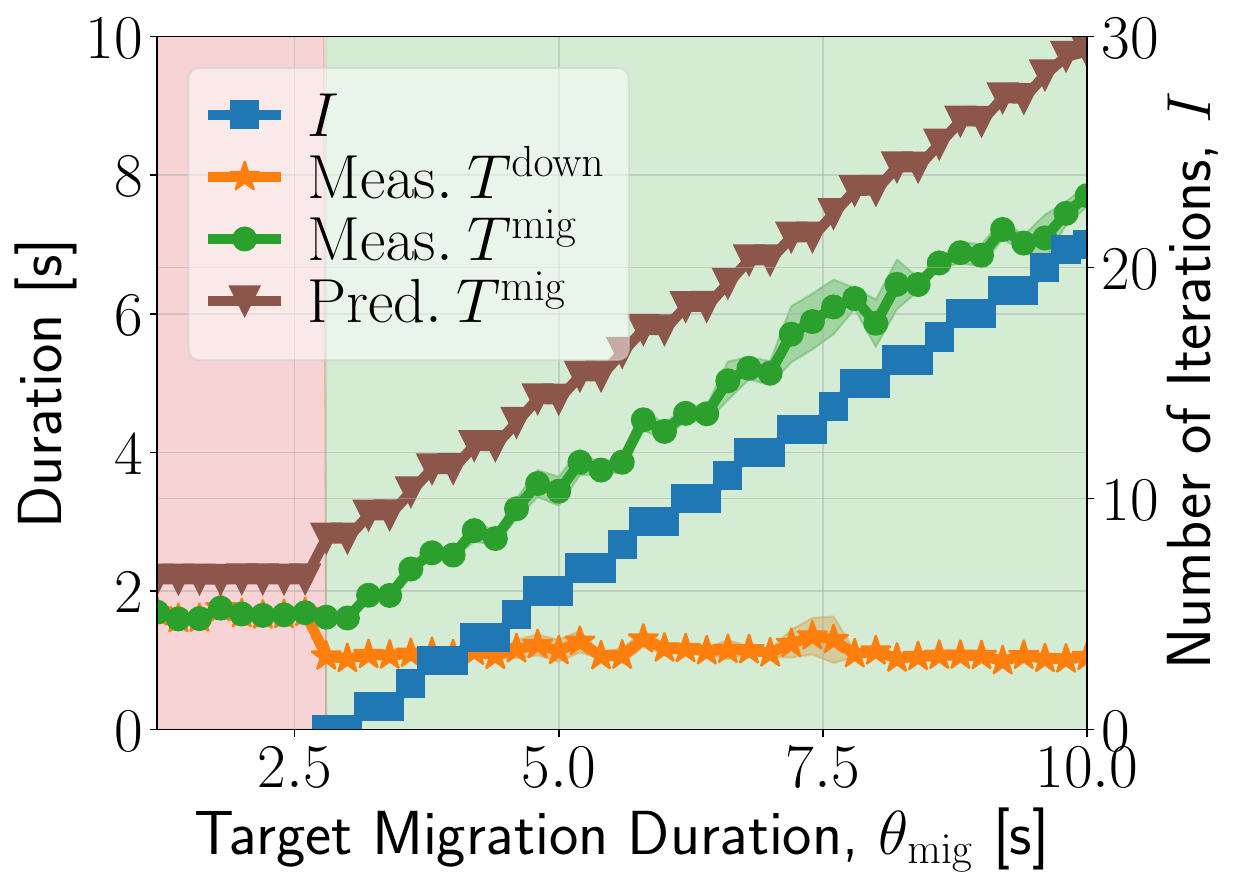}}}
  \\
  \subfloat[MOSE-MR]{\label{work05:fig:uav_min_res}{\includegraphics[width=0.7\linewidth]{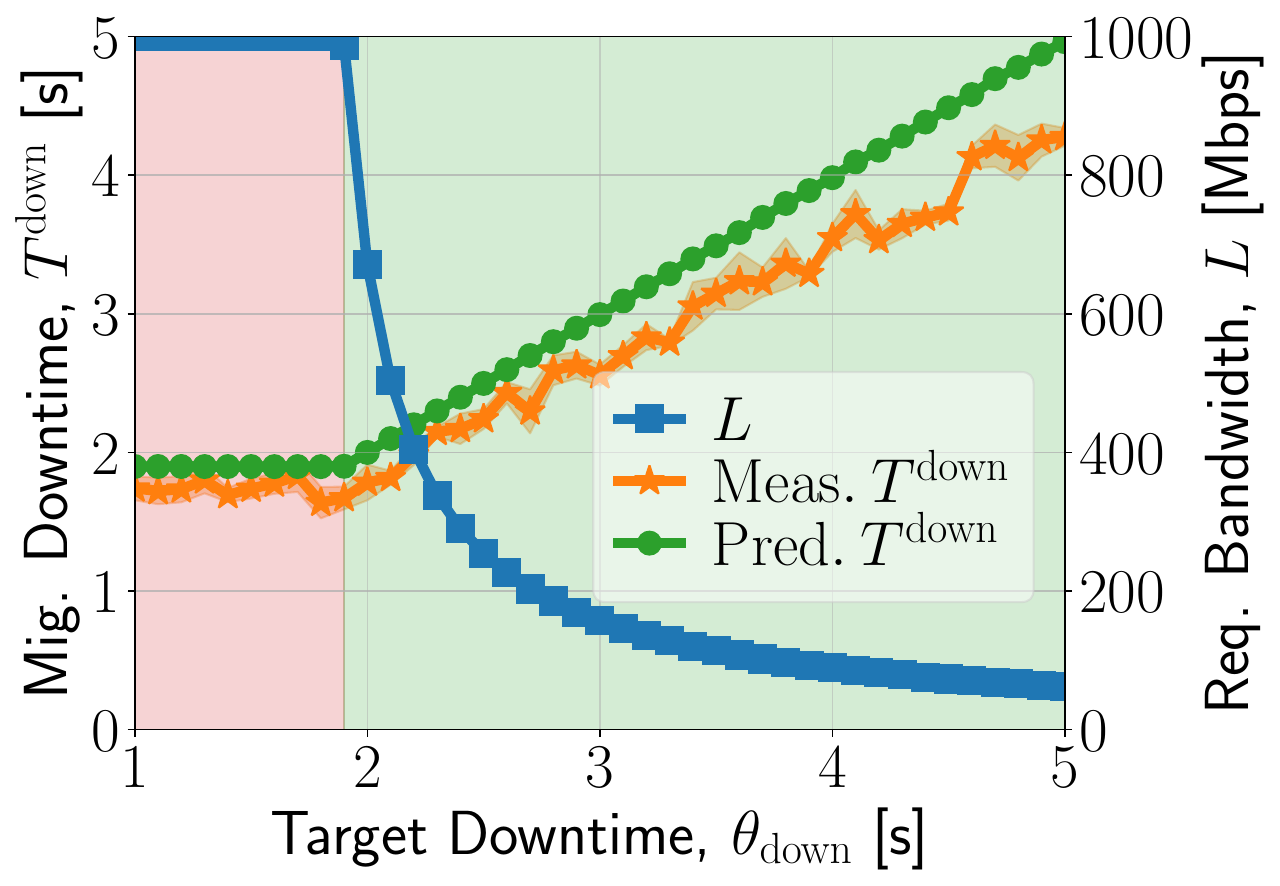}}}
  \caption{MOSE migration results for UAV autopilot MS (a) for downtime minimization (MOSE-MD) and (b) for resource usage minimization  (MOSE-MR) and for varying values of target KPIs.}
  \label{work05:fig:uav_mose_results}
  \vspace{-5mm}
\end{figure}

\begin{table*}[tb]
    \centering
    \caption{Comparison between SotA and MOSE: UAV autopilot MS migration configuration and KPIs} 
    \begin{tabular}{c|c|c|c|c|c|c|}
    \cline{2-7}
         &  & \makecell{Bandwidth\\$L$ [Mbps]} & \makecell{No. of Iterations\\$I$} & \makecell{Downtime\\$T^{\text{down}}$ [s]} & \makecell{Mig. Duration\\$T^{\text{mig}}$ [s]}  & \makecell{UAV Trajectory\\Error [m] }\\ \hline
         
        \multicolumn{1}{|c|}{\multirow{2}{*}{\textbf{{State of the Art}}}}  & Cold Migration & $1000$ & - & $3.72 \pm 0.03$ & $3.72 \pm 0.03$ & $19.36 \pm 6.39$\\ 
        
        \multicolumn{1}{|c|}{\textbf{}} & Iterative Precopy & $1000$ & $1$ & $3.62 \pm 0.04$ & $8.64 \pm 0.07$ & $18.06 \pm 7.39$\\\hline

        \multicolumn{1}{|c|}{\multirow{2}{*}{\textbf{{MOSE-MR}}}}  & $\theta_{\text{down}} = 3\,\text{s}$ & $332$ & - & $2.14 \pm 0.17$ & $2.14 \pm 0.17$ & $11.97 \pm 5.22$ \\
        
        \multicolumn{1}{|c|}{\textbf{}} & $\theta_{\text{down}} = 5\,\text{s}$ & $110$ & - & $3.63 \pm 0.29$ & $3.63 \pm 0.29$ & $17.13 \pm 4.05$ \\ \hline
        
        \multicolumn{1}{|c|}{\multirow{2}{*}{\textbf{{MOSE-MD}}}}  & $\theta_{\text{mig}} = 5\,\text{s}$ & $1000$ & $6 $ & $1.08 \pm 0.05$ & $3.7 \pm 0.10$ & $0.65 \pm 0.08$\\ 
        
        \multicolumn{1}{|c|}{\textbf{}} & $\theta_{\text{mig}} = 10\text{s}$ & $1000$ & $20 $ & $1.22 \pm 0.04$ & $8.33 \pm 0.18$ & $0.80 \pm 0.07$ \\
        \hline
    \end{tabular}
    \label{work05:tab:sota_uav_comparison}
    \vspace{-3mm}
\end{table*}

\section{MOSE Framework Exploitation}\label{work05:sec:mose_exploitation}
To demonstrate the benefits of our solution, we now demonstrate how MOSE can be used to implement and configure the stateful migration process in two practical scenarios featuring real-world MSs with diverse workload complexities. We first consider an autopilot MS controlling UAVs that provide connectivity to users in a geographical area (Sec.~\ref{work05:sub:autopilot_exploitation}). Then we focus on the migration of an ML task leveraging a UAV as data source and state-of-the-art computer vision models of different sizes to execute multi-object tracking (Sec.~\ref{work05:sub:yolo_exploitation}).

\subsection{UAV autopilot migration}\label{work05:sub:autopilot_exploitation}
In this case, we use MOSE to implement and configure stateful migration of a UAV autopilot MS residing at the network edge. After introducing our reference scenario and the testbed setup we developed, we show how MOSE allows preserving the tracking state upon migration while attaining the target migration KPIs and the vertical's objective. To conclude, we compare our solution to  state-of-the-art alternatives.

\textbf{Reference scenario and testbed setup.}
As depicted in Fig.~\ref{work05:fig:autopilot_scenario}, our reference scenario features a flying UAV and an autopilot MS that controls its trajectory. Due to the UAV's limited computational resources, the autopilot MS resides at the network edge. It periodically collects the UAV's position  and outputs the target velocity vector the UAV should enact to accurately follow the imposed trajectory. Further, since the UAV moves across the network and connects to different gNBs, we consider an edge server devoted to monitoring and managing the edge system, which  migrates the autopilot MS to the nearest edge server, i.e., the one co-located with the gNB to which the UAV is connected,  to minimize the experienced latency. We thus consider stateful migration as the key technology to address this mobility challenge.

To implement this scenario, we used the same testbed we used to validate the MOSE framework (see Sec.~\ref{work05:sub:testbed}), and we developed an autopilot MS that controls the flight of the UAV along a specific trajectory. To do so, our MS periodically gathers the UAV's GPS position and produces the target velocity the UAV should enact onto each of the 3D spatial axis. 
We consider a maximum speed of 5\,m/s, and a predefined trajectory, i.e., a circle of diameter equal to 100\,m. This trajectory is the simplest, yet most effective, to observe the trajectory error due to the migration process. We remark that this approach can be easily extended to a full-fledged autopilot that intelligently determines and updates the trajectory that the UAV should follow. 
Finally, to realistically simulate the behavior of the UAV, we use the well-known off-the-shelf PX4 framework~\cite{px4} and Gazebo simulator~\cite{gazebo}, which enable, respectively, a robust flight control and an accurate simulation of a real-world environment. Leveraging our MOSE profiling module, we characterized the  autopilot MS in terms of state size and dirty-page rate, which resulted to be  equal to 25\,MB and 0.03, respectively.

\textbf{Migration configuration.} 
Leveraging our testbed, we now show the configuration output by the MOSE orchestrator and the values obtained for the fundamental migration KPIs upon  migration of the autopilot MS. To do so, we consider the two MOSE variants previously introduced, depending on the driving vertical's objective: MOSE-MD and MOSE-MR. When the goal is to minimize the downtime, we vary the value of target migration duration (Fig.~\ref{work05:fig:uav_min_down}); when, instead, we aim at minimizing the resource usage, we vary the value of  target downtime (Fig.~\ref{work05:fig:uav_min_res}). Importantly, given the target values, we identify two main regions where the given target is met (green) or not (red).

In the case of minimizing migration downtime, 
Fig.~\ref{work05:fig:uav_min_down} presents the configured number of iterations $I$ and the fundamental KPIs $T^{\text{down}}$ and $T^{\text{mig}}$ as functions of the target migration duration $\theta^{\text{mig}}$. Specifically, for $T^{\text{mig}}$, we report both the predicted value and the one we actually measured. When $\theta^{\text{mig}}$ cannot be met (red region), MOSE-MD configures the migration process according to the Cold migration strategy. Conversely, in the green region, MOSE-MD selects Iterative PreCopy strategy and computes the value of $I$ that fulfills the given target. As expected, $\theta^{\text{mig}}$ is positively correlated with $I$ since, the larger the target KPI values, the higher the number of iterations that turns out to be feasible. Also, as the value of $I$ grows, the measured migration downtime $T^{\text{down}}$ decreases, which is consistent with the given vertical's objective, i.e., downtime minimization. Importantly,  the predicted value of $T^{\text{mig}}$, computed by MOSE-MD using our PAM model, is always larger than the measured one, confirming that PAM provides an accurate upper bound on this KPI. 

For the case of resource usage minimization,
Fig.~\ref{work05:fig:uav_min_res} shows the migration downtime $T^{\text{down}}$ and the value of network bandwidth $L$ configured by MOSE-MR, for varying values of target migration downtime $\theta^{\text{down}}$. We recall that MOSE-MR configures the migration process according to the Cold migration strategy, under which the two fundamental KPIs $T^{\text{down}}$ and $T^{\text{mig}}$ coincide.  MOSE-MR then computes the minimum amount of network bandwidth $L$ between source and destination hosts that meets the given target (green region). When instead  $\theta^{\text{down}}$ cannot be met (red region), MOSE-MR configures the migration process to use the maximum available bandwidth, i.e., 1\,Gbps. In fact,  as $\theta^{\text{down}}$ increases, the value of $L$ decreases, yielding a reduced usage of network resources. Further, the measured downtime $T^{\text{down}}$ is always shorter than $\theta^{\text{down}}$, which demonstrates PAM model's ability to provide a tight  upper bound on this KPI.

\textbf{Comparison with the state of the art.}
We now compare MOSE  with the state-of-the-art (SotA) approaches to stateful container migration. Regarding the SotA, we consider both Cold migration and Iterative PreCopy strategy, and fix realistic values for both bandwidth $L$ and number of PreCopy iterations $I$. We recall that our work is the first that  accurately configures the migration process by leveraging an analytical model. Further, we assess MOSE performance under its two variants, MOSE-MD and MOSE-MR, and for varying values of the target KPI. Specifically, we set $\theta^{\text{mig}}$ equal to 5 and 10 seconds for the former and $\theta^{\text{down}}$ equal to 3 and 5 seconds for the latter. Table~\ref{work05:tab:sota_uav_comparison} shows, for each scenario, the migration configuration output by MOSE along with the measured values for the fundamental KPIs and the UAV trajectory error caused by the migration process. Notably, compared to SotA Cold Migration, MOSE-MR yields up to 89\% reduction in the allocated network bandwidth, with negligible impact on the migration KPIs and the UAV trajectory error. Similarly, when the goal is to minimize the downtime, MOSE-MD, compared to SotA Iterative PreCopy, substantially reduces both downtime and UAV trajectory error, up to {70\%} and {97\%}, respectively, while exhibiting a comparable value of migration duration.
{\em Such significant reduction demonstrates MOSE's effectiveness in attaining migration of time-critical MSs, such as the UAV autopilot we considered, for which the UAV trajectory error can be lowered up to {0.65\,m} -- a   negligible error} relatively  to the 100\,m diameter of the area we considered.

\subsection{Multi-object tracking task migration}\label{work05:sub:yolo_exploitation}
Here we use MOSE to implement and configure stateful migration of a machine-learning (ML) task. We start by introducing our reference scenario, featuring an edge-assisted UAV and a Multi-Object Tracking (MOT) application. Then, using our testbed, we demonstrate how MOSE outputs an accurate migration configuration that attains both target migration KPIs and the vertical's objective for varying ML model complexity.
Finally, we present a probabilistic characterization of MOSE orchestration features, and a comparison with the migration strategies in prior art.

\begin{figure}[tb]
  \centering
  \includegraphics[width=\linewidth]{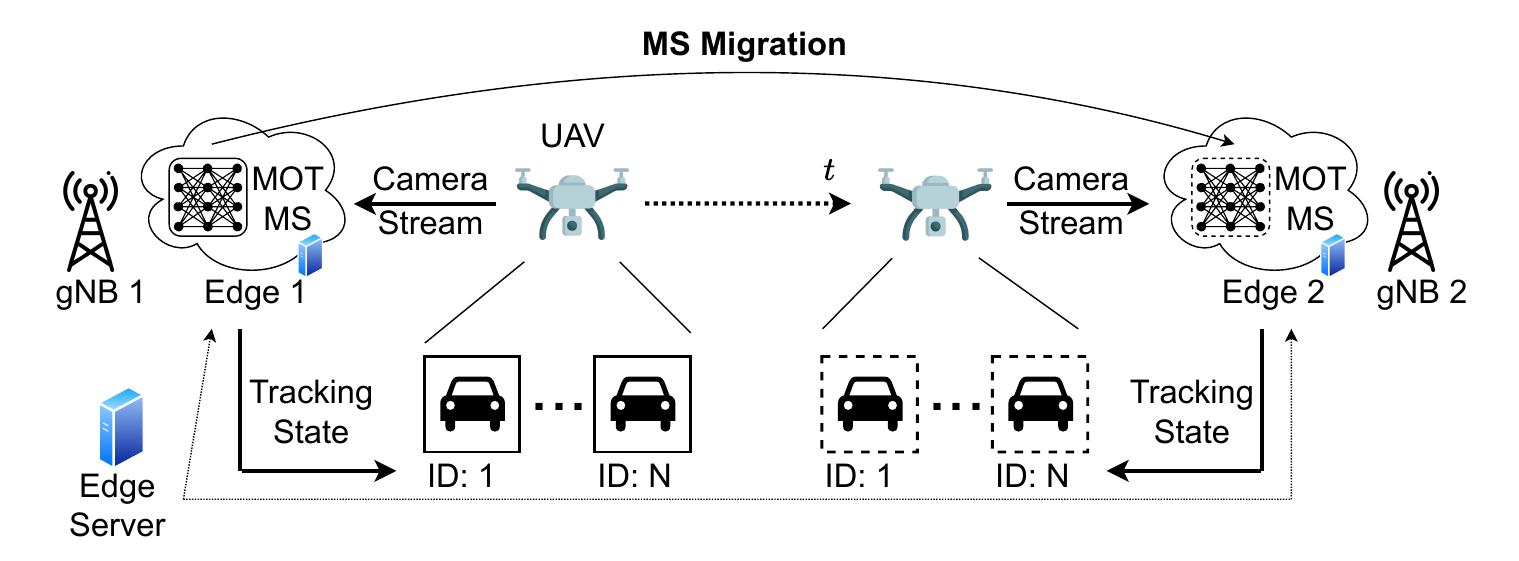}
  \caption{Reference scenario for the migration of the machine-learning task for multi-object tracking.}
  \label{work05:fig:yolo_scenario}
  \vspace{-4mm}
\end{figure}

\textbf{Reference scenario and testbed setup.} We focus on the practical scenario depicted in Fig.~\ref{work05:fig:yolo_scenario}, featuring a UAV with an on-board camera and a MOT task residing at the network edge that performs inference on a camera stream. As the UAV moves across the considered area, it  connects to different gNBs. Due to the UAV's limited computational resources, the MOT task must be deployed at the edge in the form of an MS. We thus consider an edge server responsible for monitoring and managing the edge system that, to minimize the experienced latency, deploys such MSs on the nearest edge server, i.e., the one co-located with the gNB the UAV is currently connected to. We thus consider stateful container migration as the key technology to address such mobility challenge while ensuring continuous proximity of the edge MSs to the UAV and preserving the tracking state.

To implement the above scenario, we employ the same testbed we developed to validate the MOSE framework (see Sec.~\ref{work05:sub:testbed}) and consider Ultralytics~\cite{yolov8_ultralytics} as the MS to be statefully migrated. 
Ultralytics is a renown off-the-shelf solution that integrates (i) YOLOv8~\cite{redmon2016yolo}, a widely used collection of pre-trained real-time object detection and image segmentation models, and (ii) BoT-SORT~\cite{aharon2022bot}, a cutting-edge multi-object tracking algorithm that, leveraging motion modeling and re-identification, assigns a unique ID to each detected object as the video progresses. Specifically, YOLOv8 consists of multiple deep-learning models, each  identified by a {\em size} indicator, from nano to extralarge, thus allowing to establish the desired trade-off between inference accuracy and latency. Indeed, as shown in~\cite{yolov8}, a higher model complexity in terms of number of parameters generally corresponds to a higher inference accuracy but also larger latency and resource consumption. We account for this trade-off by reporting MOSE configuration results for all the available models. Interestingly, {\em MOSE can be proactively used to select which model size best fits the desired targets and user's QoE level.} 
Further, to mimic the behavior of a UAV streaming an on-board camera feed, we use MediaMTX~\cite{mediamtx}, a real-time media server supporting a wide variety of video codecs and streaming protocols. For our experiments, we deploy MediaMTX in VM3, which, we recall, acts as an end device, and use it to generate, under default settings, an RTSP video stream. Such stream is then consumed by the Ultralytics MS to attain the MOT task.

\begin{figure}[tb]
  \centering
  \includegraphics[width=0.8\linewidth]{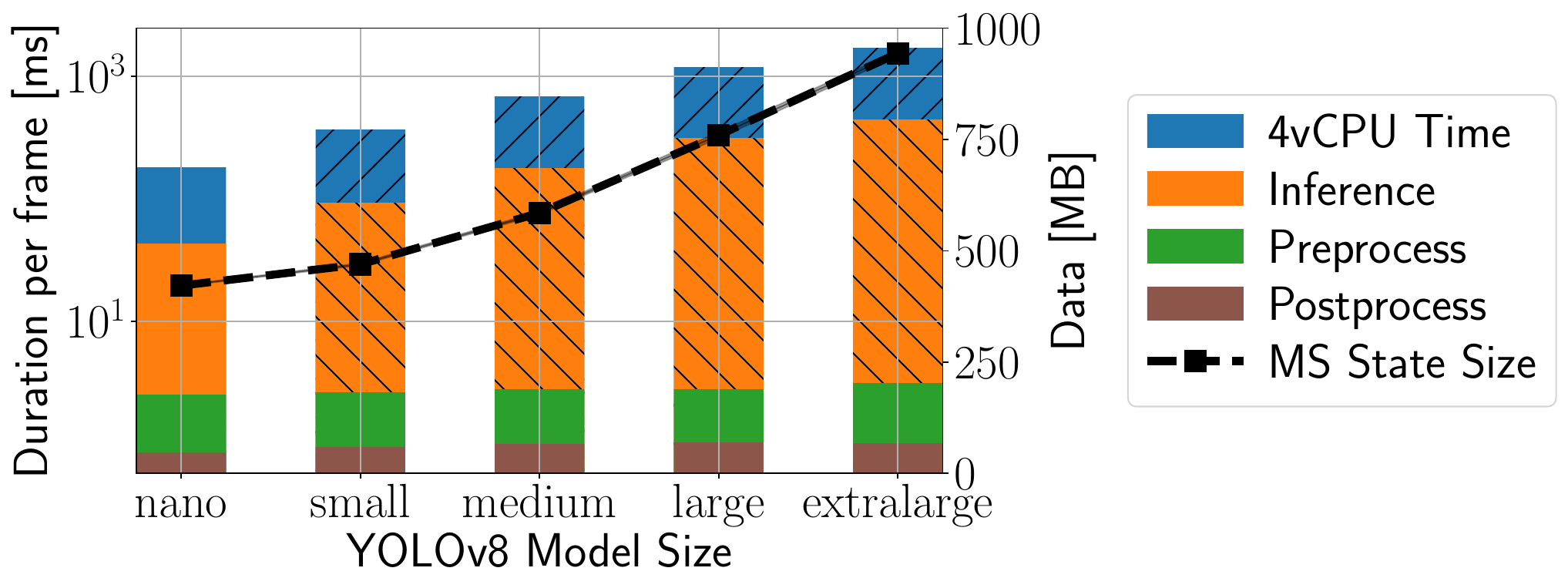}
  \caption{Ultralytics MS performance vs.\,YOLO model size.}
  \label{work05:fig:yolo_perf}
  \vspace{-3mm}
\end{figure}

Fig.~\ref{work05:fig:yolo_perf} presents a preliminary characterization of the Ultralytics MS performance as a function of the YOLO model size. It shows the total CPU time per frame, accounting for the 4\,vCPUs allocated, and the related internal contributions, i.e., the time the ML model needs to provide an inference result and the time needed to pre-process and post-process the input video frame. These results can be used to derive (i) the inference rate $\rho$, i.e., the number of video frames that our testbed is capable of processing per second, and (ii) the number of frames lost due to the migration downtime, derived as $\rho T^{\text{down}}$. With regard to the experimental results presented in the following, we show the migration downtime as the fundamental KPI, considering that this can be easily mapped onto  frame loss as an alternative reference metric.
Further, leveraging our MOSE profiling module (see Sec.~\ref{work05:sub:mose_agent}), we characterize the Ultralytics MS in terms of state size, which, as highlighted in Fig.~\ref{work05:fig:yolo_perf}, is proportional to the configured YOLO model size, and the normalized dirty-page rate, which, instead, resulted to be approximately equal to 0.66, independently of the YOLO model size.

\begin{figure*}[htp]
    \centering
    \subfloat[$\theta_{\text{mig}} = 25\,\text{s}$]{\label{work05:fig:yolo_num_iter_25}{\includegraphics[width=0.25\textwidth]{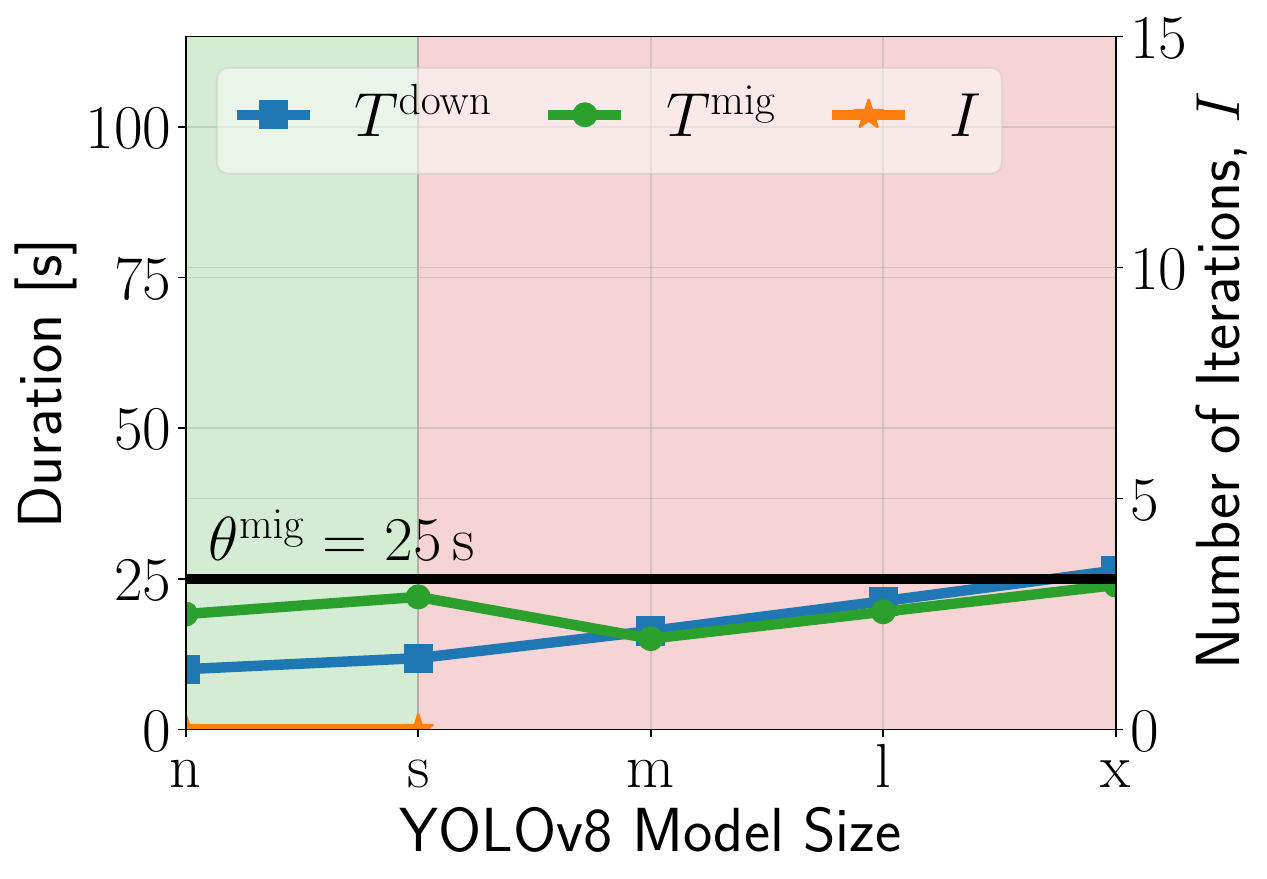}}}
    \subfloat[$\theta_{\text{mig}} = 50\,\text{s}$]{\label{work05:fig:yolo_num_iter_50}{\includegraphics[width=0.25\textwidth]{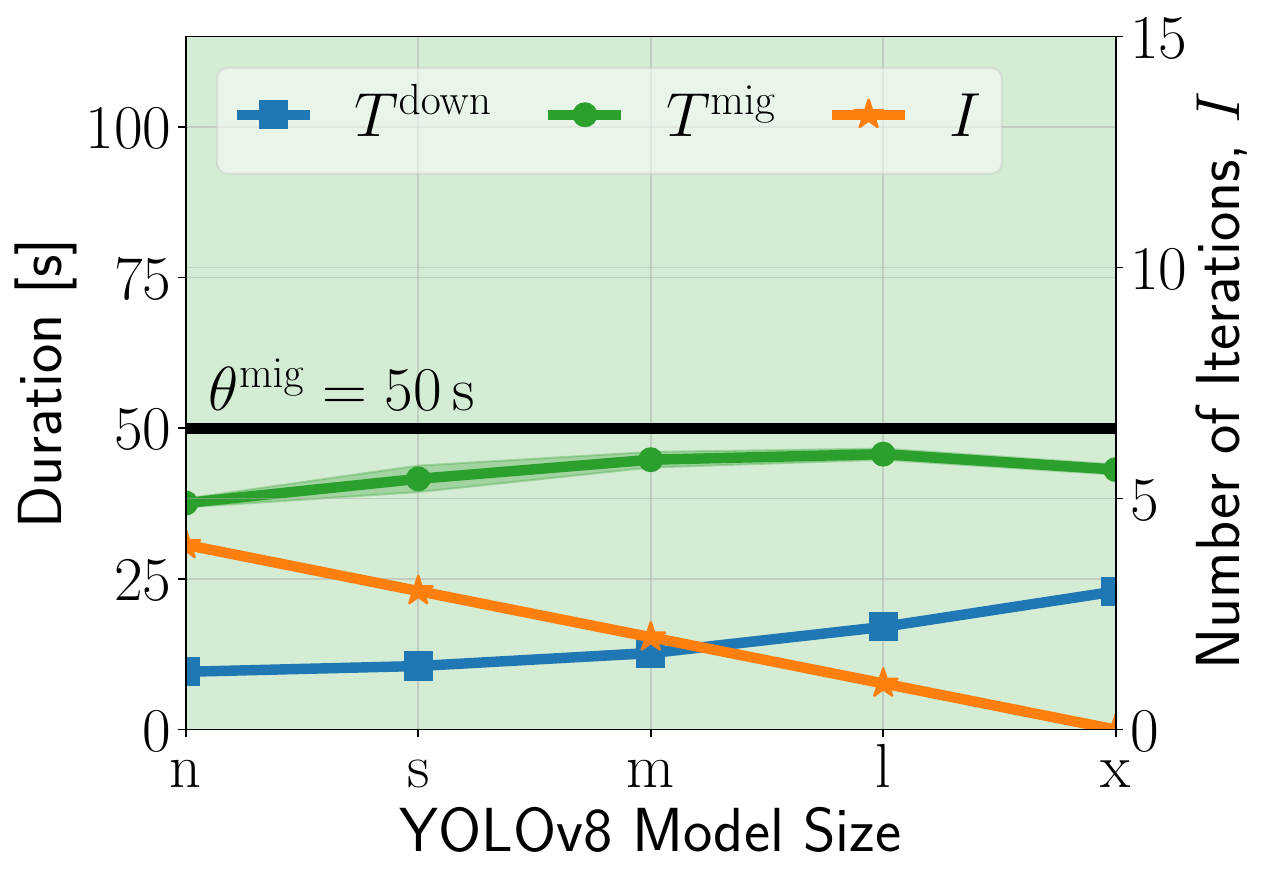}}}
    \subfloat[$\theta_{\text{mig}} = 75\,\text{s}$]{\label{work05:fig:yolo_num_iter_75}{\includegraphics[width=0.25\textwidth]{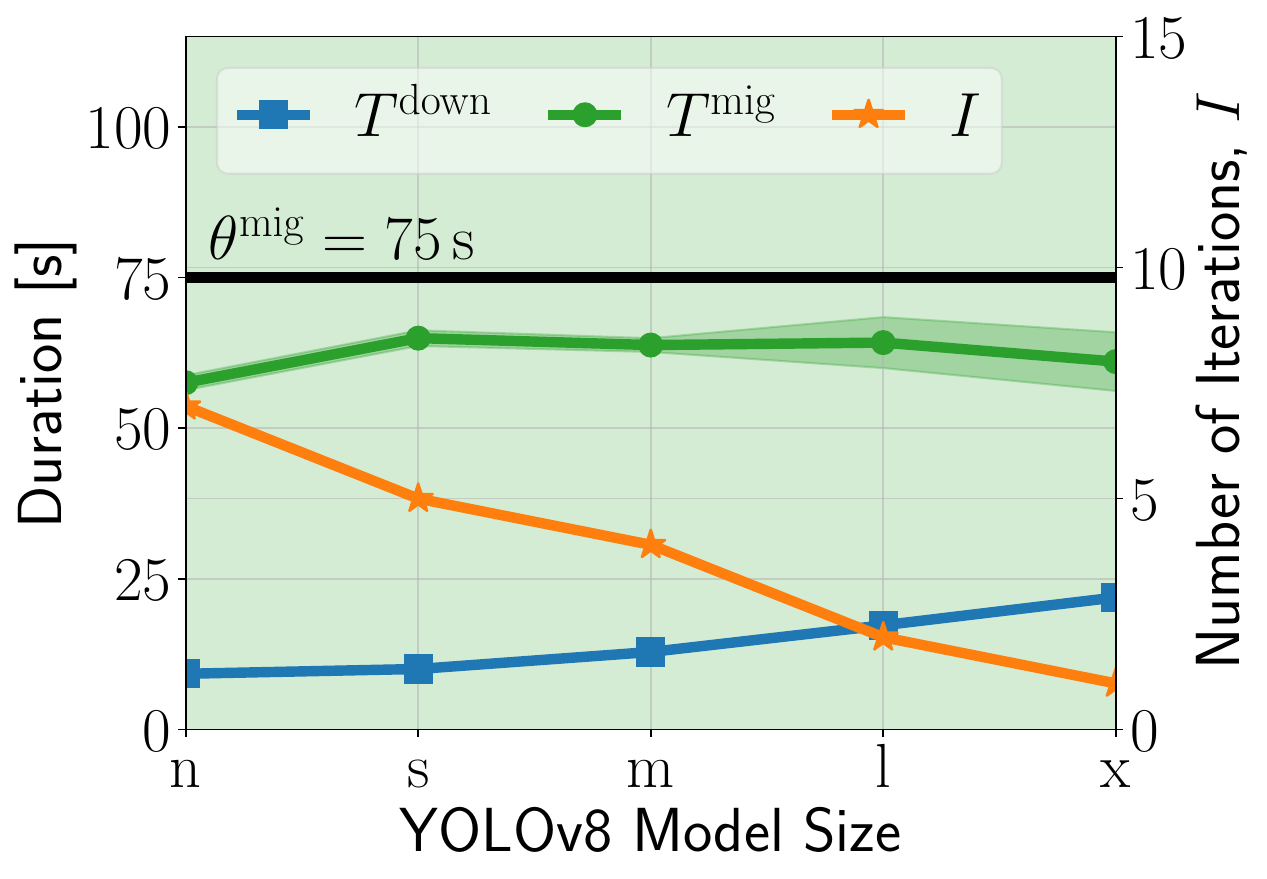}}}
    \subfloat[$\theta_{\text{mig}} = 100\,\text{s}$]{\label{work05:fig:yolo_num_iter_100}{\includegraphics[width=0.25\textwidth]{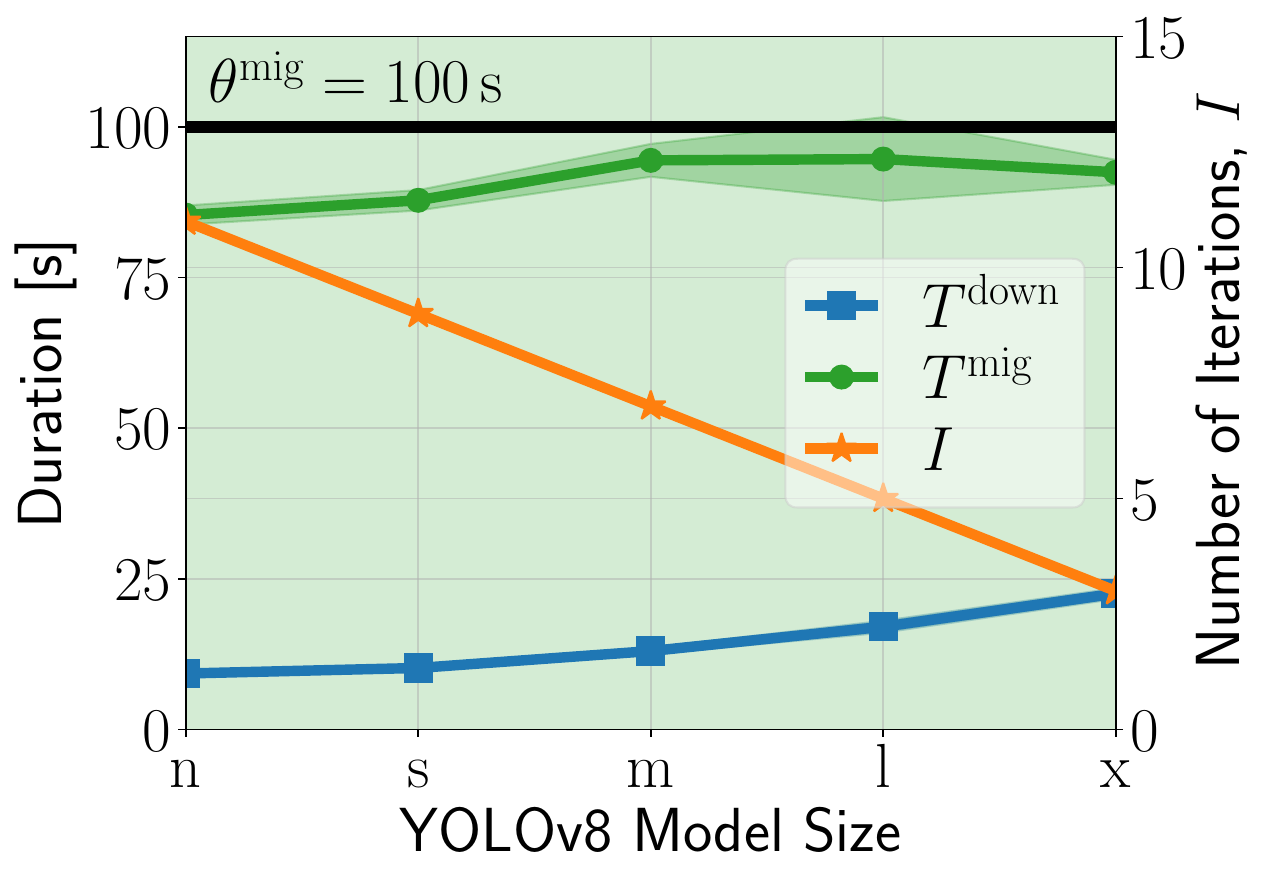}}}
    \caption{MOSE-MD performance and configuration for varying target migration duration and YOLO model size.}
    \label{work05:fig:yolo_min_down}
    \vspace{-3mm}
\end{figure*}

\begin{figure*}[htp]
    \centering
    \subfloat[$\theta_{\text{down}} = 15\,\text{s}$]{\label{work05:fig:yolo_req_band_15}{\includegraphics[width=0.25\textwidth]{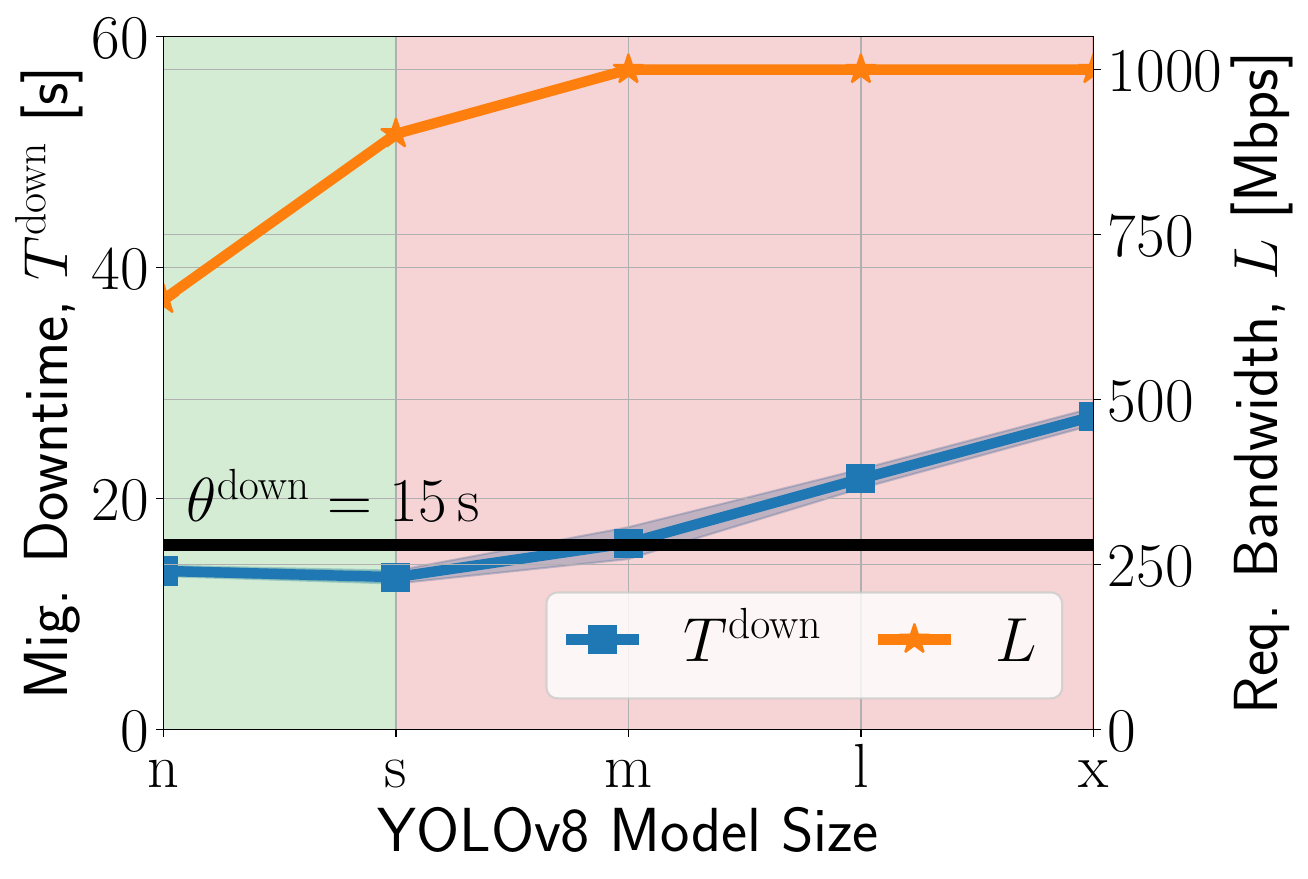}}}
    \subfloat[$\theta_{\text{down}} = 30\,\text{s}$]{\label{work05:fig:yolo_req_band_30}{\includegraphics[width=0.25\textwidth]{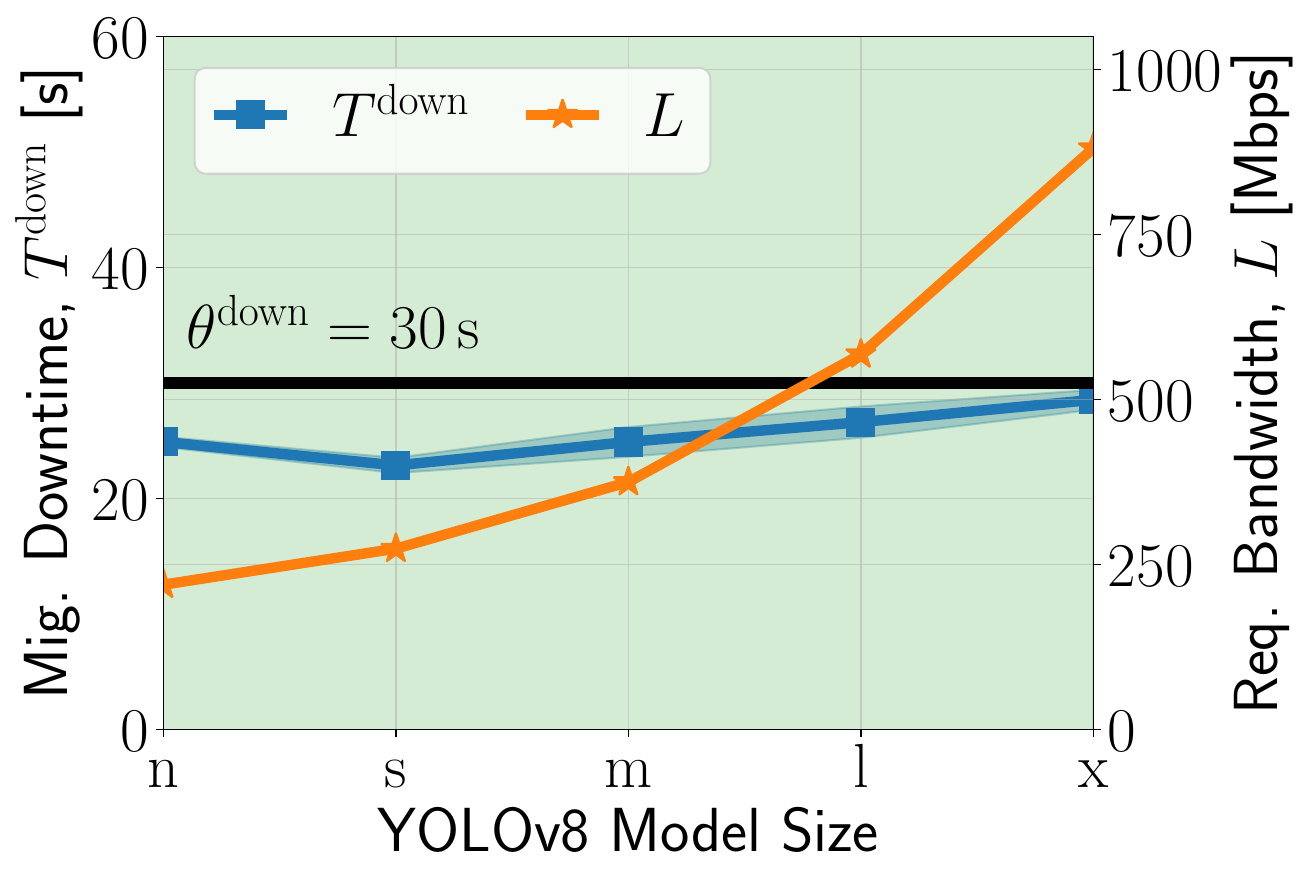}}}
    \subfloat[$\theta_{\text{down}} = 45\,\text{s}$]{\label{work05:fig:yolo_req_band_45}{\includegraphics[width=0.25\textwidth]{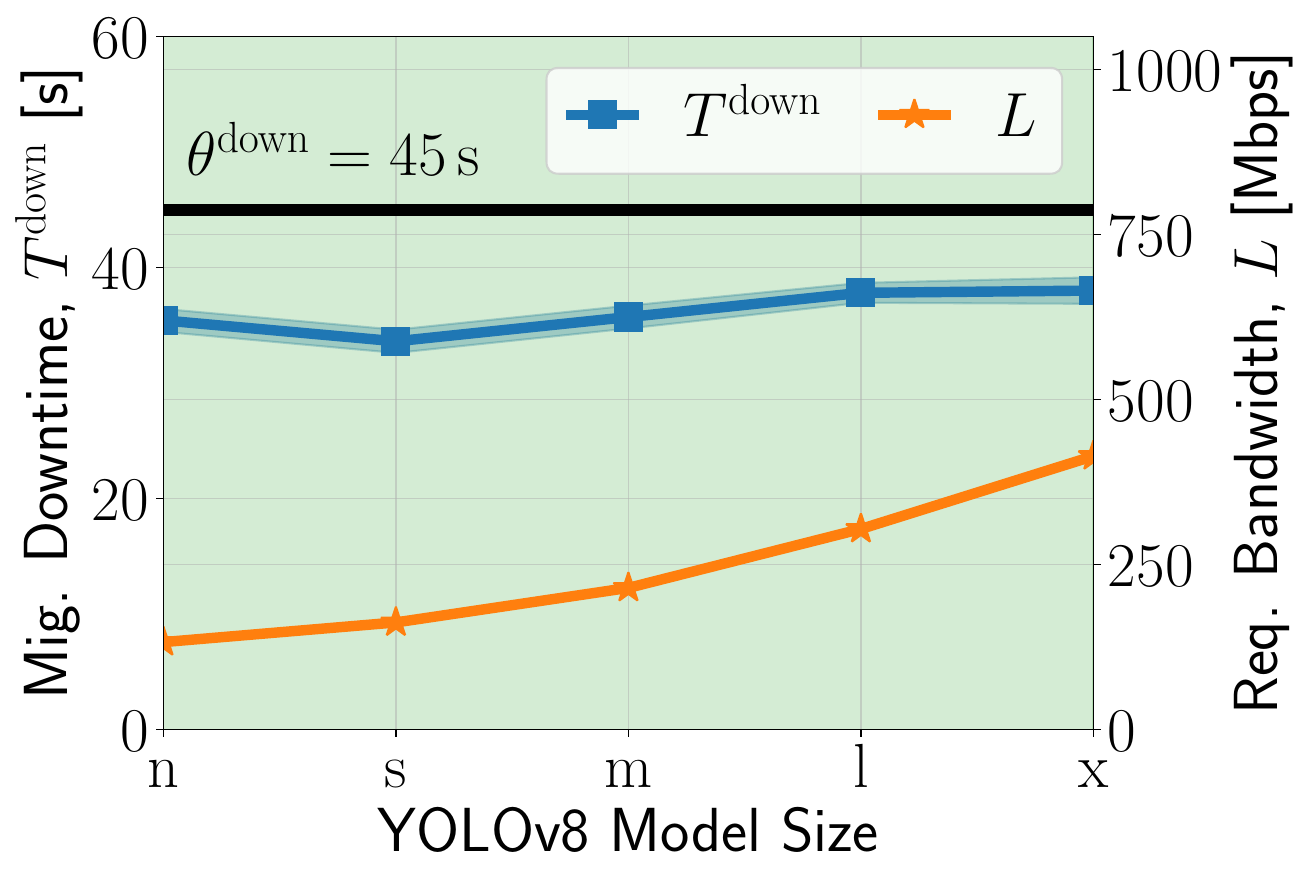}}}
    \subfloat[$\theta_{\text{down}} = 60\,\text{s}$]{\label{work05:fig:yolo_req_band_60}{\includegraphics[width=0.25\textwidth]{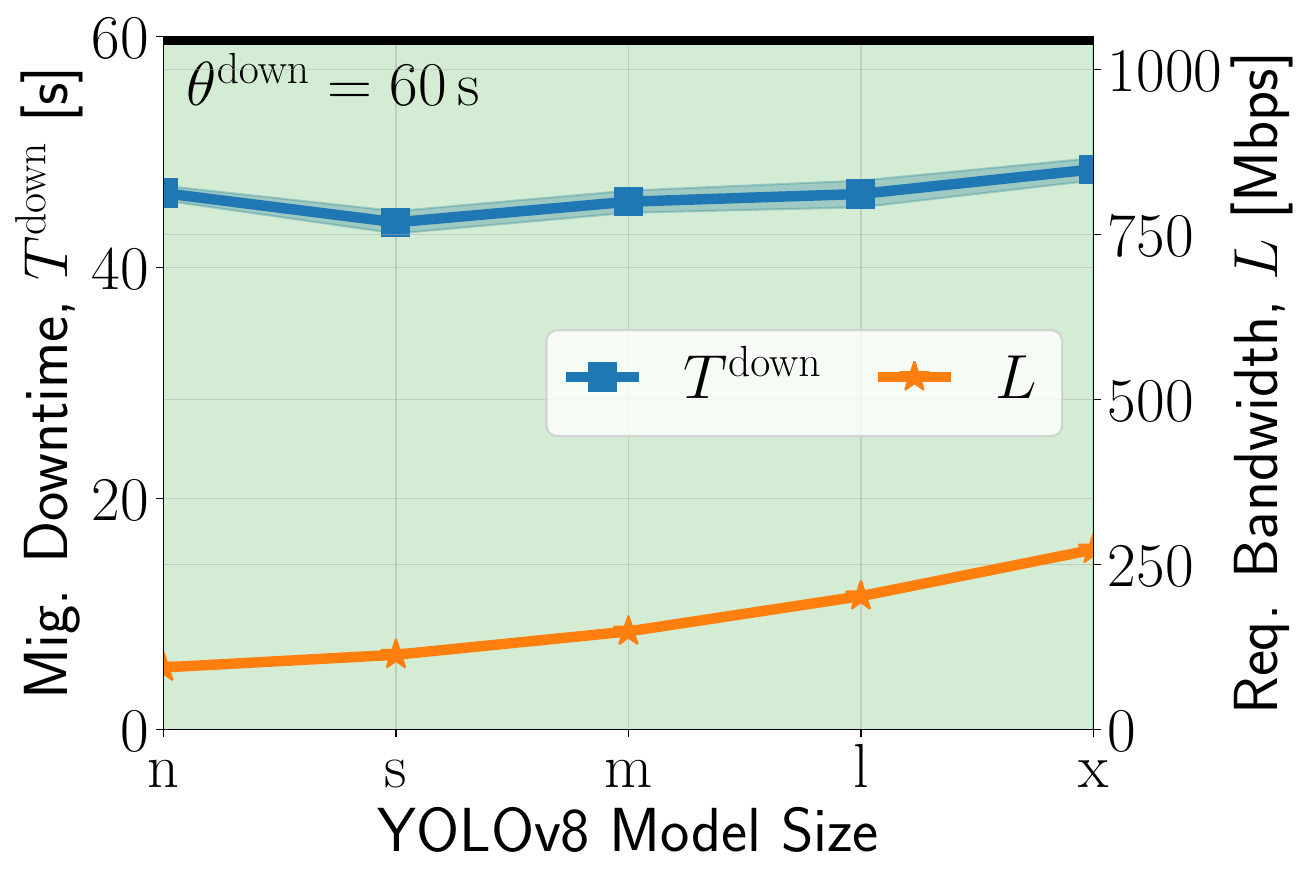}}}
    \caption{MOSE-MR performance and configuration for varying target downtime and YOLO model size.}
    \label{work05:fig:yolo_min_res}
\end{figure*}

\begin{table*}[!htb]
    \centering
    \caption{Comparison between MOSE and SotA: MOT MS migration configuration and KPIs}
    \begin{tabular}{c|c|c|c|c|c|}
    \cline{2-6}
         &  & \makecell{Bandwidth\\$L$ [Mbps]} & \makecell{No. of Iterations\\$I$} & \makecell{Downtime\\$T^{\text{down}}$ [s]} & \makecell{Mig. Duration\\$T^{\text{mig}}$ [s]}  \\ \hline
        \multicolumn{1}{|c|}{\multirow{2}{*}{\textbf{{State of the Art}}}}  & Cold Migration & $1000$ & - & $11.83 \pm 0.16$ & $11.83 \pm 0.16$ \\ 
        \multicolumn{1}{|c|}{\textbf{}} & Iterative Precopy & $1000$ & $15$ & $10.40 \pm 0.14$ & $134.17 \pm 1.37$ \\
        \hline
        \multicolumn{1}{|c|}{\multirow{2}{*}{\textbf{{MOSE-MR}}}}  & $\theta_{\text{down}} = 15\,\text{s}$ & $651.60 \pm 25.18$ & - & $13.62 \pm 0.35$ & $13.62 \pm 0.35$ \\ 
        \multicolumn{1}{|c|}{\textbf{}} & $\theta_{\text{down}} = 60\,\text{s}$ & $94.20 \pm 1.32$ & - & $46.33 \pm 0.54$ & $46.33 \pm 0.54$ \\
        \hline
        \multicolumn{1}{|c|}{\multirow{2}{*}{\textbf{{MOSE-MD}}}}  & $\theta_{\text{mig}} = 20\,\text{s}$ & $1000$ & $0$ & $9.80 \pm 0.22$ & $18.87 \pm 0.51$ \\ 
        \multicolumn{1}{|c|}{\textbf{}} & $\theta_{\text{mig}} = 100\text{s}$ & $1000$ & $11$ & $9.25 \pm 0.26$ & $85.39 \pm 1.33$ \\
        \hline
    \end{tabular}
    \label{work05:tab:sota_yolo_comparison}
    \vspace{-3mm}
\end{table*}

\textbf{Migration configuration.}
Leveraging our testbed, we now show the configuration output by the orchestrator and the values obtained for the fundamental migration KPIs upon MOT MS migration. To do so, we consider the two MOSE variants previously introduced, depending on the considered vertical's objective: MOSE-MD and MOSE-MR. When the goal is to minimize the downtime, we vary the target migration duration (Fig.~\ref{work05:fig:yolo_min_down}); when, instead, we aim at minimizing the resource usage, we vary the target downtime (Fig.~\ref{work05:fig:yolo_min_res}). Again, given  the target values, we identify two main regions where the target is met (green) or not (red).

For the case of  migration downtime minimization,
Fig.~\ref{work05:fig:yolo_min_down} depicts the fundamental KPIs $T^{\text{down}}$ and $T^{\text{mig}}$ and the configured number of PreCopy iterations $I$ as the YOLO model size and the values of target migration duration $\theta^{\text{mig}}$ vary. When $\theta^{\text{mig}}$ cannot be met (red region), MOSE-MD configures the migration process according to the Cold migration strategy. Otherwise (green region), MOSE-MD selects Iterative PreCopy strategy and computes the number of iterations $I$ that fulfills the target. As the model size increases, the value of $I$ decreases, which is due to the increased MS state size. Further, the experienced downtime $T^{\text{down}}$ slightly increases with the model size but, as $\theta^{\text{mig}}$ varies, it remains almost constant, regardless of the value of $I$. This is due to the fact that, as described above, the MOT MS  is characterized by a high dirty-page rate, which  reduces the effectiveness of Iterative PreCopy strategy in minimizing the downtime in a way that is proportional to $I$. Notably, the measured migration duration $T^{\text{mig}}$ is always shorter than $\theta^{\text{mig}}$, underlining that the PAM model provides an accurate upper bound on this KPI. 

In the case of minimizing resource usage, Fig.~\ref{work05:fig:yolo_min_res} presents the migration downtime $T^{\text{down}}$ and the required bandwidth $L$ for different values of  YOLO model size and  target downtime $\theta^{\text{down}}$. We recall that, to attain this goal, MOSE-MR configures the migration process according to Cold migration strategy and, in such a case, downtime and migration duration coincide. Moreover, when $\theta^{\text{down}}$ cannot be met (red region), MOSE-MR configures the migration to use the maximum available bandwidth, i.e., 1\,Gbps. Otherwise (green region), it computes the minimum amount of network bandwidth between source and destination hosts that meets the corresponding target. Accordingly, the value of $L$ is positively correlated with the model size, and  increases  with $\theta^{\text{down}}$. Importantly, the actually experienced downtime $T^{\text{down}}$ is always shorter than $\theta^{\text{down}}$.

\textbf{Comparison with the state of the art.}
We now present a comparison of MOSE performance with the  SotA approaches  under the MOT MS use case. Specifically, we report the migration configuration and the experienced KPI values for multiple scenarios, i.e., (i) SotA, when either Cold migration or Iterative PreCopy is used, (ii) MOSE, for varying vertical's objective, i.e., MOSE-MD and MOSE-MR variants, and (iii) for each objective, two representative values of target KPI. When SotA is considered, the  bandwidth $L$ and the number of iterations $I$ are fixed a priori; the former is set to  the maximum value, i.e., 1\,Gbps, and the latter to 15 since, as shown in previous works, e.g., \cite{ibrahim2011optimized}, the larger the number of iterations, the lower the downtime. Instead, as mentioned, our solution is the first that does not  rely on fixed parameters, rather, leveraging an analytical model, it accurately configures and orchestrates the migration process.

The results are summarized in Table~\ref{work05:tab:sota_yolo_comparison}. We recall that, when the goal is to minimize resource usage, MOSE-MR selects Cold migration strategy and computes the value of $L$ that fulfills the target downtime $\theta_{\text{down}}$.
Remarkably, MOSE-MR allows for up to 91\% reduction of  allocated bandwidth, thus reducing network congestion and preventing resource over-provisioning.
Similarly, when the goal is to minimize the migration downtime, MOSE-MD selects Iterative PreCopy and computes the value of $I$ that allows  meeting the target migration duration $\theta_{\text{mig}}$.
Differently from SotA, MOSE-MD, by accounting for the target KPI value, outputs a migration configuration that yields a comparable downtime and, concurrently, up to 86\% reduction of the observed migration duration, hence a lower consumption of computing resources.

\begin{figure*}[htp]
    \centering
    \subfloat[nano]{\label{work05:fig:yolo_pmf_n}{\includegraphics[width=0.2\textwidth]{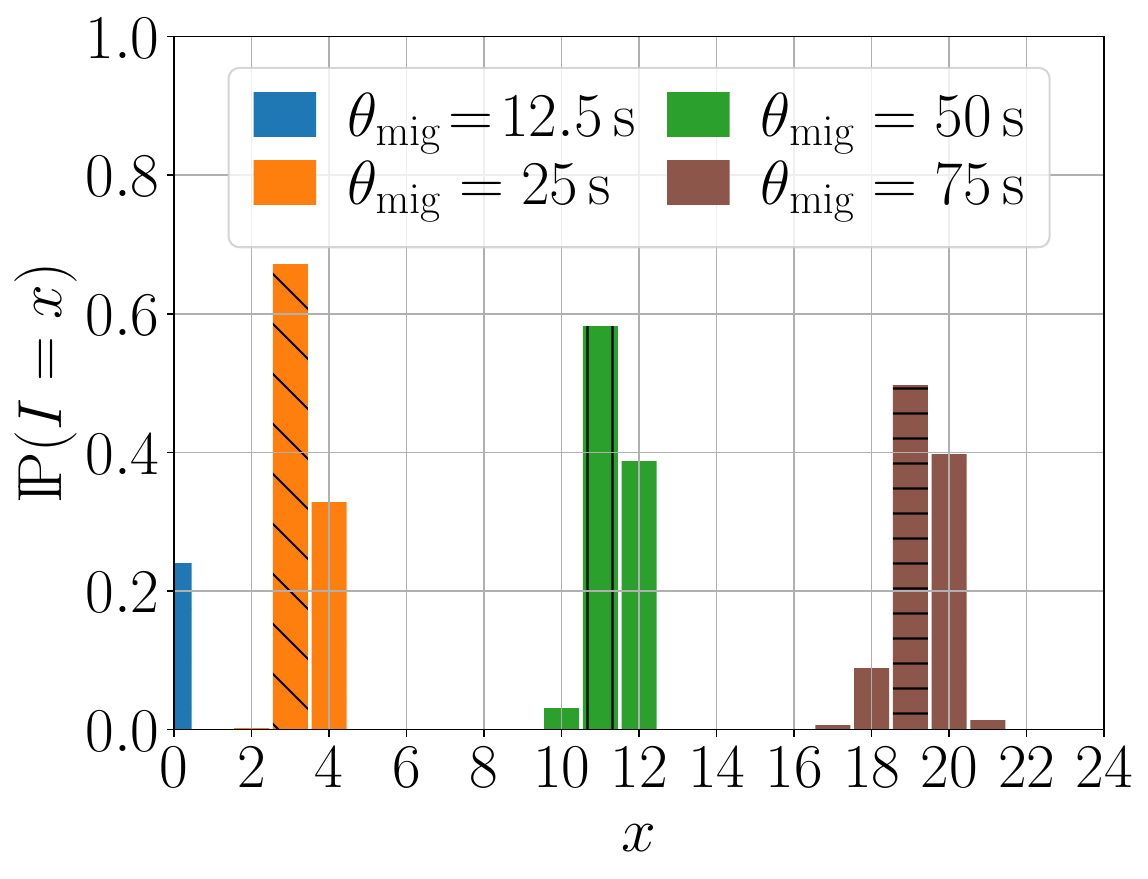}}}
    \subfloat[small]{\label{work05:fig:yolo_pmf_s}{\includegraphics[width=0.2\textwidth]{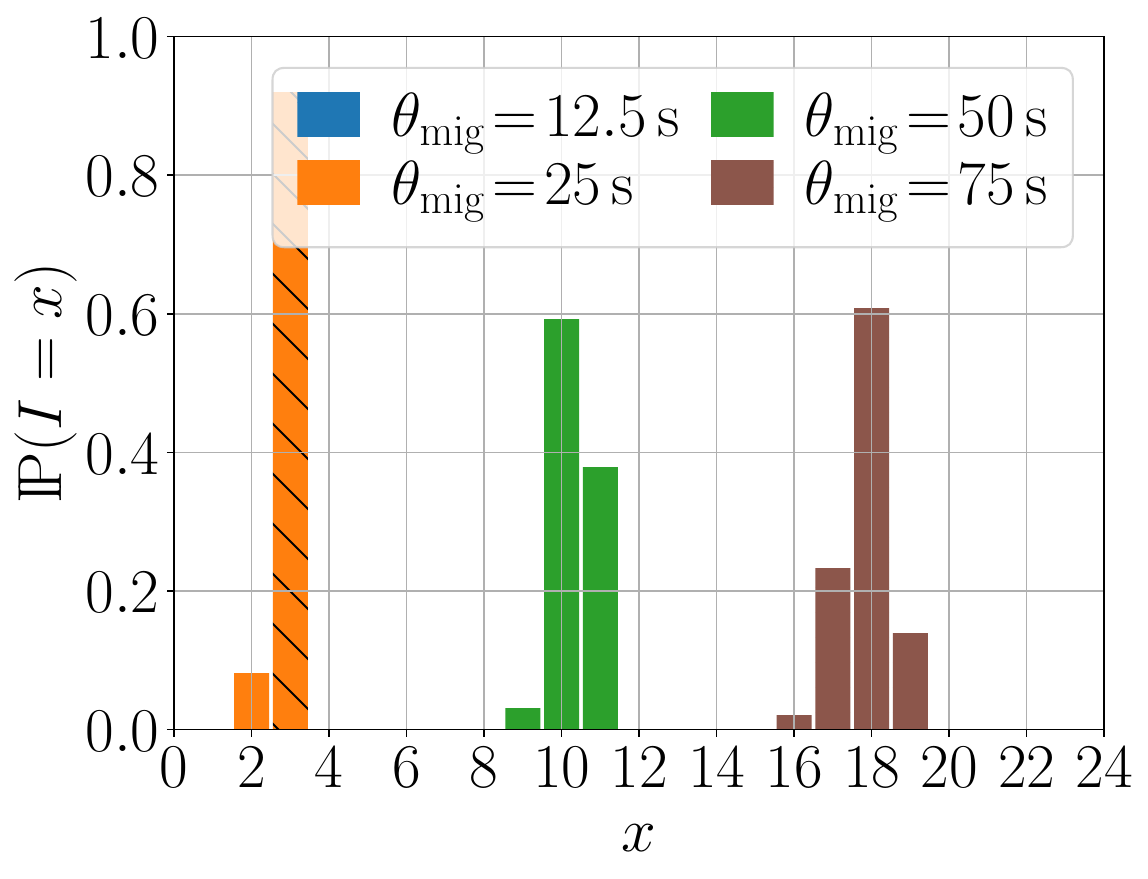}}}
    \subfloat[medium]{\label{work05:fig:yolo_pmf_m}{\includegraphics[width=0.2\textwidth]{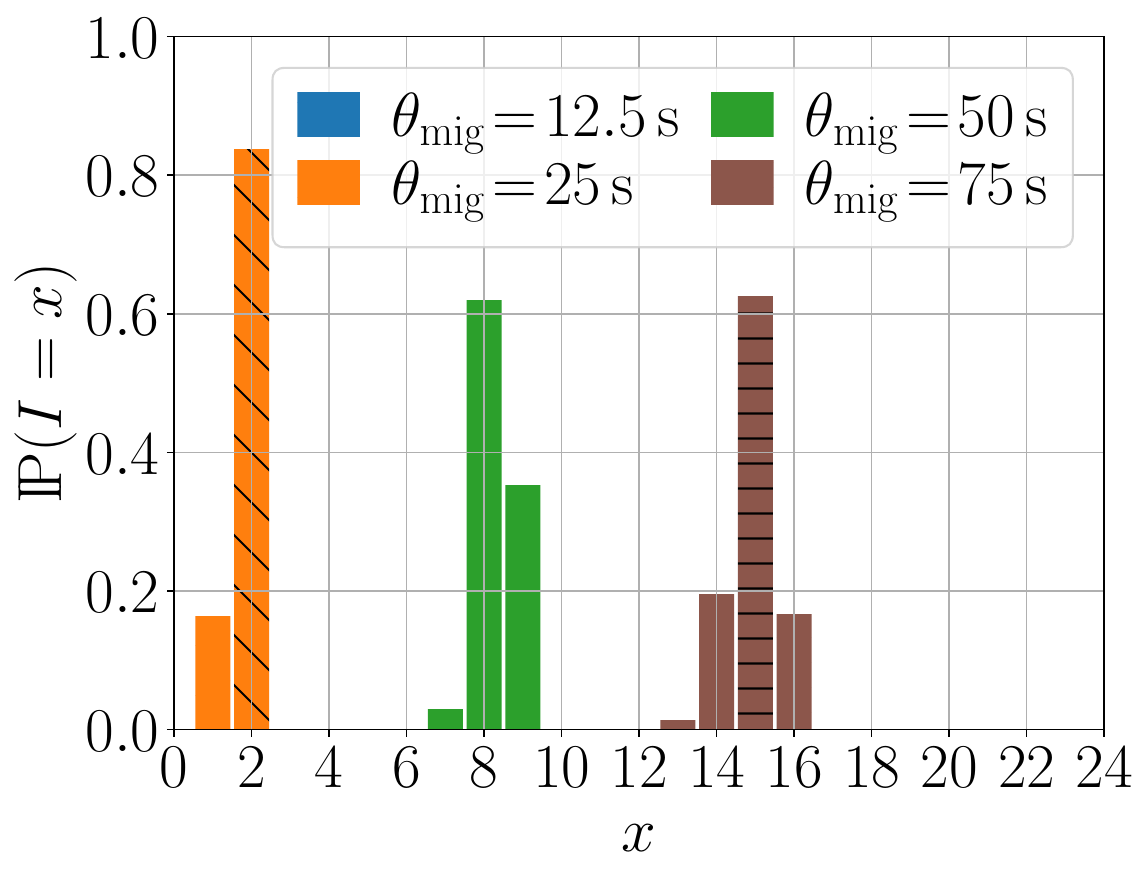}}}
    \subfloat[large]{\label{work05:fig:yolo_pmf_l}{\includegraphics[width=0.2\textwidth]{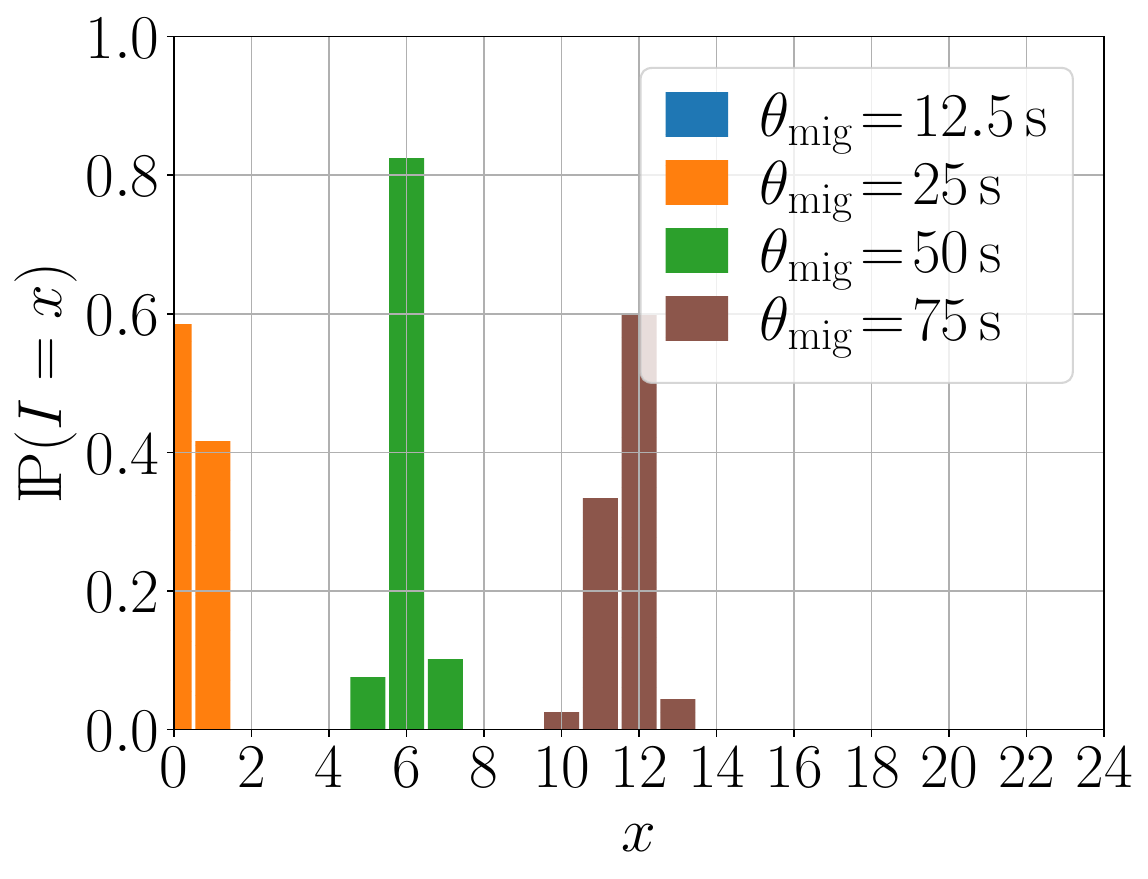}}}
    \subfloat[extralarge]{\label{work05:fig:yolo_pmf_x}{\includegraphics[width=0.2\textwidth]{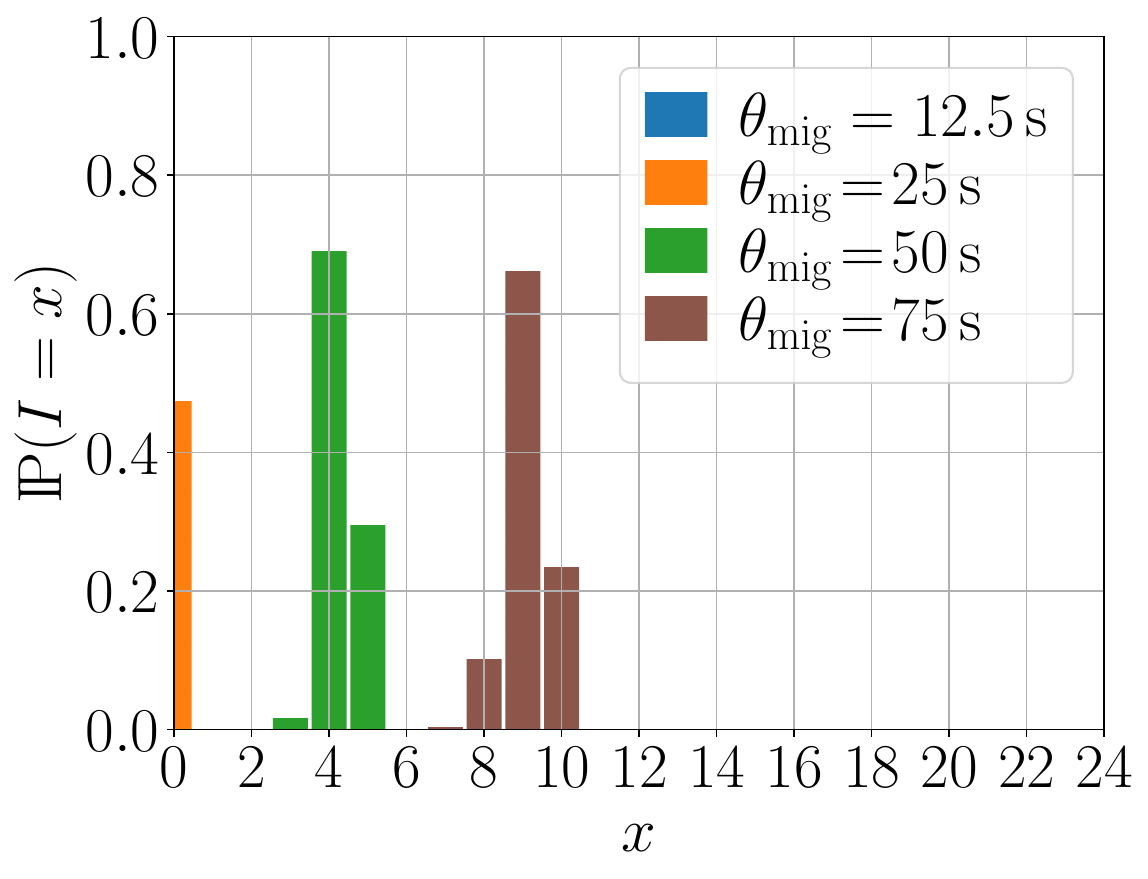}}}
    \caption{MOSE-MD probability mass function for varying YOLO model size and target migration duration.}
    \label{work05:fig:yolo_pmf}
    \vspace{-3mm}
\end{figure*}

\begin{figure*}[htp]
    \centering
    \includegraphics[width=\textwidth]{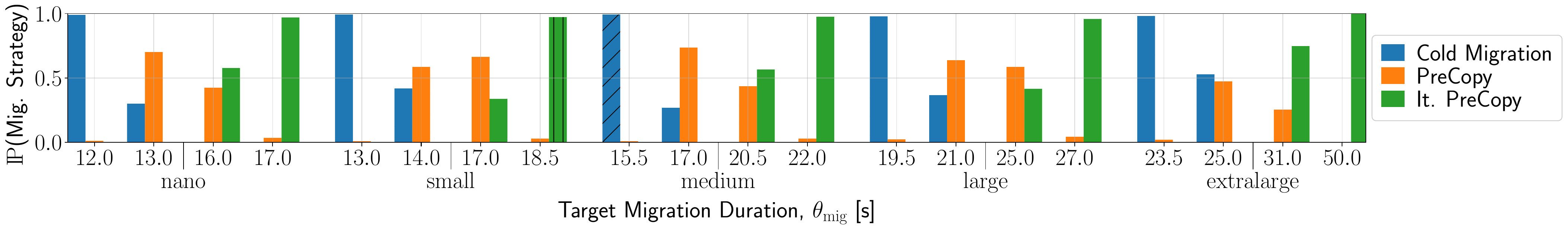}
    \caption{MOSE migration strategy probability for downtime minimization and for varying target migration duration and YOLO model size.}
    \label{work05:fig:pam_strategy_prob}
    \vspace{-5mm}
\end{figure*}

\textbf{Probabilistic analysis.}
To thoroughly characterize our Migration Designer module (see Sec.~\ref{work05:sub:mose_orchestrator}), we present a probabilistic analysis of MOSE orchestration features. To do so, we consider the maximum available bandwidth for the link between source and destination edge hosts as a truncated, normally-distributed random variable with 1\,Gbps as mean value and 100\,Mbps as standard deviation, and observe the migration configuration output by MOSE. Importantly, we focus on the more complex vertical's objective to minimize the migration downtime. Indeed, while for resource usage minimization MOSE-MR always selects Cold migration strategy, when the goal is to attain downtime minimization, MOSE-MD selects the strategy that best matches the KPI targets.

Fig.~\ref{work05:fig:yolo_pmf} depicts the probability mass function of $I$ versus the target migration duration $\theta^{\text{mig}}$ and for varying YOLO model sizes. Notably, as the value of $\theta^{\text{mig}}$ increases, the range of possible values for $I$ increases. Further, while a tight target like 12.5\,s can only be achieved for the nano model size and with a particularly small probability, a loose target like 75\,s can always be met, regardless of the chosen YOLO model.

 Fig.~\ref{work05:fig:pam_strategy_prob} shows the probability that MOSE-MD selects a  migration strategy out of Cold migration, PreCopy, and Iterative PreCopy. Such probability is evaluated as a function of the target migration duration $\theta^{\text{mig}}$ and for varying YOLO model sizes. Specifically, for each model size, four relevant values of $\theta^{\text{mig}}$ are considered. The first is a value below which Cold migration is always selected, and, similarly, for the fourth value and larger ones, Iterative PreCopy is always feasible. Instead, the second and third values show the probability of PreCopy being feasible and thus selected by MOSE-MD as the most appropriate strategy.

{\em Summary.} Our results show that MOSE: (i) effectively attains migration of ML tasks with varying complexity and accounting for both KPI target values and different vertical's objectives; (ii)  greatly outperforms state-of-the-art approaches, yielding a migration configuration that prevents resource over-provisioning; (iii) can be used to proactively identify the most appropriate KPI target values and model complexity in the case of randomly distributed system parameters.

\section{Conclusions}\label{work05:sec:conclusions}
Stateful MS migration has emerged as the fundamental technology to support service mobility at the network edge. Nevertheless, several technical challenges related to the implementation and orchestration of the migration process still need to be addressed.
To fill such gaps, we envisioned MOSE, a novel framework that efficiently implements stateful MS migration at the edge and effectively orchestrates the migration process to ensure a minimal impact on the user's QoE.
Leveraging our testbed and realistic MSs, we showed that MOSE greatly outperforms the state of the art, with a reduction of the service downtime up to {77\%}. Then, we validated MOSE for varying migration KPI targets and accounting for different vertical's objectives, thus proving that MOSE can effectively address scenarios where reliability and robustness are of primary importance, e.g., those featuring time-critical applications.
To further demonstrate the effectiveness of our solution, we considered two relevant practical use cases featuring, respectively, a UAV autopilot MS, and a multi-object tracking MS. In both cases, MOSE effectively attains migration of such MSs while greatly outperforming the state-of-the-art approaches by up to {97\%} reduction of the UAV trajectory error, for the former, and up to 91\% reduction of network resources utilization, for the latter.

\bibliographystyle{IEEEtran}
\bibliography{myBibliography.bib}

\vspace{-1cm}
\begin{IEEEbiographynophoto}
{Antonio Calagna} is a post-doc researcher at Politecnico di Torino, from which he received his Ph.D. degree in  2025.
\end{IEEEbiographynophoto}
\vskip -2.5\baselineskip plus -1fil
\begin{IEEEbiographynophoto}
{Yenchia Yu} is a Ph.D. student at Politecnico di Torino, from which he received his M.Sc. degree in 2022.
\end{IEEEbiographynophoto}
\vskip -2.5\baselineskip plus -1fil
\begin{IEEEbiographynophoto}
{Paolo Giaccone} received his Ph.D. degree from the Politecnico di Torino, Italy in 2001, where he is currently a Full Professor.
\end{IEEEbiographynophoto}
\vskip -2.5\baselineskip plus -1fil
\begin{IEEEbiographynophoto}
{Carla Fabiana Chiasserini} is Full Professor with the Politecnico di Torino, Italy, and a Research Associate with CNR and CNIT.
\end{IEEEbiographynophoto}

\end{document}